\global\def\draftcontrol{0}

   \def\versionno{ qgpmagnetic}

\catcode`\@=11

\expandafter\ifx\csname draftcontrol\endcsname\relax\global\def\draftcontrol{0}
\fi

{\count255=\time\divide\count255 by 60
\xdef\hourmin{\number\count255}
\multiply\count255 by-60\advance\count255 by\time
\xdef\hourmin{\hourmin:\ifnum\count255<10 0\fi\the\count255}}
\def\draftdate{\number\month/\number\day/\number\year\ \ \ \hourmin }

\newcommand\makepapertitle{\par
  \begingroup
    \renewcommand\thefootnote{\@fnsymbol\c@footnote}%
    \def\@makefnmark{\rlap{\@textsuperscript{\normalfont\@thefnmark}}}%
    \long\def\@makefntext##1{\parindent 1em\noindent
            \hb@xt@1.8em{%
                \hss\@textsuperscript{\normalfont\@thefnmark}}##1}%
     \newpage
     \global\@topnum\z@   
     \@makepapertitle
     \thispagestyle{empty}\@thanks
  \endgroup
  \setcounter{footnote}{0}%
  \global\let\thanks\relax
  \global\let\makepapertitle\relax
  \global\let\@makepapertitle\relax
  \global\let\@thanks\@empty
  \global\let\@author\@empty
  \global\let\@date\@empty
  \global\let\@title\@empty
  \global\let\title\relax
  \global\let\author\relax
  \global\let\date\relax
  \global\let\and\relax
  \def\version{\let\version\@version\@gobble}
}
\def\@makepapertitle{%
  \newpage
   \ifnum\draftcontrol=1 {}
   \version\versionno
   \vskip 3em%
   \else
   \hfill\hbox to 3cm {\parbox{4cm}{\@pubnum}\hss}%
   \vskip 3em%
   \fi
   \begin{center}%
   \let \footnote \thanks
     {\LARGE {\@title}}%
     \vskip 1.5em%
     {\normalsize
       \lineskip .5em%
       \begin{tabular}[t]{c}%
         \@author
       \end{tabular}\par}%
     \vskip 1.5em%
     {\@bstract}%
     \end{center}%
     \vskip 1.5em
     \@date%
   \par
}

\gdef\@pubnum{}
\def\pubnum#1{%
  \gdef\@pubnum{#1}}

\gdef\@bstract{}
\def\Abstract#1{%
  \gdef\@bstract{%
   \parbox{\textwidth-0pc}{%
   \centerline{\bf Abstract}\penalty1000%
\kern.2cm%
\noindent
\renewcommand\baselinestretch{1.0}%
{#1}}}
}

\def\ps@paper{\let\@mkboth\@gobbletwo%
     \ifnum\draftcontrol=1
    \def\@oddfoot{\hbox to \textwidth{\tiny \versionno \hfil\tiny\draftdate}%
    \hskip -\textwidth \hbox to \textwidth{\hfil\rm\thepage\hfil}}%
     \else\def\@oddfoot{\hbox to \textwidth{\hfil\rm\thepage\hfil}}
     \fi
     \let\@evenfoot\@oddfoot
}

\def\body{\clearpage
          \pagestyle{paper}
    }

\def\@version#1{\ifnum\draftcontrol=1
\typeout{}\typeout{#1}\typeout{}
\vskip3mm\centerline{\hbox{\fbox{\normalsize{\tt DRAFT -- #1 -- }
                   {\draftdate}}}}\vskip3mm
\fi}
\let\version\@version
\long\def\eqlabel#1{\ifnum\draftcontrol=1
                    \tag@false  
                    \tag*{(\theequation) \hbox to -0.2cm{\hspace{0cm}\small{#1}\hss}}
                    \refstepcounter{equation}
                    \edef\@currentlabel{\theequation}
                    \ltx@label{#1}          
                    \else
                    \label{#1}
                    \fi
                    }
\let\st@bibitem\@bibitem
\let\st@lbibitem\@lbibitem
\ifnum\draftcontrol=1
  \def\@bibitem#1{%
    \st@bibitem{#1}\a@@label{#1}\ignorespaces}
  \def\@lbibitem[#1]#2{%
    \st@lbibitem[#1]{#2}\a@@label{#2}\ignorespaces}
  \def\a@@label#1{%
    \gdef\a@lab{\smash{\normalfont\small#1}}
    \ifvmode
      \if@inlabel
        \global\setbox\@labels\hbox{%
          \llap{\a@lab\let\a@lab\relax
                \kern\@totalleftmargin\kern\marginparsep}%
          \box\@labels}%
      \fi
    \fi}
\fi

\documentclass[12pt,letterpaper]{article}

\usepackage{amsmath,amssymb,array,calc,epsfig,rotating,bm}
\usepackage[sort]{cite}
\usepackage{graphicx}
\usepackage{psfrag,verbatim}
\usepackage{xcolor}

\ifnum\draftcontrol=1
\fi

\renewcommand\baselinestretch{1.25}
\renewcommand\section{\@startsection {section}{1}{\z@}%
                                   {-3.5ex \@plus -1ex \@minus -.2ex}%
                                   {2.3ex \@plus.2ex}%
                                   {\normalfont\large\bfseries}}
\renewcommand\subsection{\@startsection{subsection}{2}{\z@}%
                                   {-3.25ex\@plus -1ex \@minus -.2ex}%
                                   {1.5ex \@plus .2ex}%
                                   {\normalfont\normalsize\bfseries}}
\renewcommand\subsubsection{\@startsection{subsubsection}{3}{\z@}%
                                   {-3.25ex\@plus -1ex \@minus -.2ex}%
                                   {1.5ex \@plus .2ex}%
                                   {\normalfont\normalsize\it}}
\renewcommand\paragraph{\@startsection{paragraph}{4}{\z@}%
                                   {-3.25ex\@plus -1ex \@minus -.2ex}%
                                   {1.5ex \@plus .2ex}%
                                   {\normalfont\normalsize\bf}}

\numberwithin{equation}{section}

\def\revise#1       {\raisebox{-0em}{\rule{3pt}{1em}}%
                     \marginpar{\raisebox{.5em}{\vrule width3pt\
                     \vrule width0pt height 0pt depth0.5em
                     \hbox to 0cm{\hspace{0cm}{%
                     \parbox[t]{4em}{\raggedright\footnotesize{#1}}}\hss}}}}

\newcommand\nxt[1]  {\\\fnxt#1}
\newcommand{\ie}{{\it i.e.,}\ }
\newcommand{\eg}{{\it e.g.,}\ }

\def\cftd {{\mathbb{CFT}_{diag}}}
\def\cftstu {{\mathbb{CFT}_{STU}}}
\def\cftpw {{\mathbb{CFT}_{PW,m=0}}}
\def\cftminfty {{\mathbb{CFT}_{PW,m=\infty}}}
\def\ncft {{n\mathbb{CFT}_m}}

\def\cale         {{\cal E}}
\def\calf         {{\cal F}}

\def\calm         {{\cal M}}
\def\caln         {{\cal N}}
\def\calo         {{\cal O}}
\def\calp         {{\cal P}}

\def\zet          {{\mathbb Z}}

\def\del          {\partial}

\def\sqr#1#2{{\vcenter{\vbox{\hrule height.#2pt
 \hbox{\vrule width.#2pt height#1pt \kern#1pt
 \vrule width.#2pt}\hrule height.#2pt}}}}

\def\hx{\hat{x}}
\def\hb{\hat{b}}
\def\hy{\hat{y}}
\def\hz{\hat{z}}

\def\aa1{\phi}
\def\cc1{\psi}

\catcode`\@=12

\begin{document}


\title{\bf QGP universality in a magnetic field?}

\date{April 3, 2020}

\author{
Alex Buchel$^{1,2,3}$ and Bruno Umbert$^{1,3}$\\[0.4cm]
\it $^1$Department of Applied Mathematics\\
\it $^2$Department of Physics and Astronomy\\ 
\it University of Western Ontario\\
\it London, Ontario N6A 5B7, Canada\\
\it $^3$Perimeter Institute for Theoretical Physics\\
\it Waterloo, Ontario N2J 2W9, Canada\\[0.4cm]
}

\Abstract{We use top-down holographic models to study the thermal equation of
state of strongly coupled quark-gluon plasma in external magnetic
field.  We identify different conformal and non-conformal theories
within consistent truncations of $\caln=8$ gauged supergravity in five
dimensions (including STU models, gauged $\caln=2^*$ theory) and show
that the ratio of the transverse to the longitudinal pressure
$P_T/P_L$ as a function of $T/\sqrt{B}$ can be collapsed to a
'universal' curve for a wide range of the adjoint hypermultiplet
masses $m$.  We stress that this does not imply any hidden
universality in magnetoresponse, as other observables do not exhibit
any universality. Instead, the observed collapse in $P_T/P_L$ is
simply due to a strong dependence of the equation of state on the
(freely adjustable) renormalization scale: in other words, it is
simply a fitting artifact. Remarkably, we do uncover a different
universality in $\caln=2^*$ gauge theory in the external magnetic
field: we show that magnetized $\caln=2^*$ plasma has a critical point
at $T_{crit}/\sqrt{B}$ which value varies by $2\%$ (or less) as
$m/\sqrt{B}\in [0,\infty)$. At criticality, and for large values of
$m/\sqrt{B}$, the effective central charge of the theory scales as
$\propto \sqrt{B}/m$.
}

\makepapertitle

\body

\version\versionno
\tableofcontents

\section{Introduction and summary}\label{intro}

In \cite{Endrodi:2018ikq} the authors used the recent lattice QCD equation of state (EOS) data in
the presence of a background magnetic field \cite{Bali:2013esa,Bali:2014kia}, and the holographic EOS
results\footnote{Studied for the first time in \cite{DHoker:2009mmn}.} for the
strongly coupled $\caln=4$ $SU(N)$ maximally supersymmetric Yang-Mills (SYM) to argue for the
universal magnetoresponse. While $\caln=4$ SYM is conformal, the scale invariance is explicitly
broken by the background magnetic field $B$ and its thermal equilibrium stress-energy tensor
is logarithmically sensitive to the choice of the renormalization scale. It
was shown in \cite{Endrodi:2018ikq} that both the QCD and the
$\caln=4$ data (with optimally adjusted renormalization scale) for the pressure anisotropy $R$,
\begin{equation}
R\equiv \frac{P_T}{P_L}\,,
\eqlabel{defr}
\end{equation}
\ie defined as a ratio of the transverse $P_T$ to the longitudinal $P_L$
pressure\footnote{We take a constant magnetic field to be  $\mathbf{B}=B \mathbf{e}_z$ so that $P_T$ and $P_L$
are correspondingly the $\langle T_{xx}\rangle=\langle T_{yy}\rangle$ and the $\langle T_{zz}\rangle$ components of the
stress-energy tensor.}, collapse onto a
single universal curve as a function of $T/\sqrt{B}$, at least for $T/\sqrt{B}\gtrsim 0.2$
or correspondingly for $R\gtrsim 0.5$, see Fig.~6 of \cite{Endrodi:2018ikq}. The authors do mention
that the 'universality' is somewhat fragile: besides the obvious fact that large-$N$ $\caln=4$ SYM
is not QCD (leading to inherent ambiguities as to how precisely one would match the renormalization
schemes in both theories --- hence the authors opted for the freely-adjustable renormalization
scale in SYM), one observes the universality in $R$, but not in other thermodynamic quantities
(\eg ${P_T}/{\cale}$ --- the ratio of the transverse pressure to the energy density). 

So, is there a universal magnetoresponse? In this paper we address this question
in a controlled setting: specifically, we consider holographic models of gauge theory/string
theory correspondence \cite{Maldacena:1997re,Aharony:1999ti} where all the four-dimensional
strongly coupled gauge theories discussed have the same ultraviolet fixed point ---
$\caln=4$ SYM. We discuss two classes of theories:
\begin{itemize}
\item conformal gauge theories corresponding to different consistent truncations of  
$\caln=8$ gauged supergravity in five dimensions\footnote{In this class of theories
there is a well motivated choice of the renormalization scale --- namely,
it is natural to have it be the same for all the theories in the class.} \cite{Bobev:2010de};
\item non-conformal $\caln=2^*$ gauge theory ($\caln=4$ SYM with a mass term for the
$\caln=2$ hypermultiplet) \cite{Pilch:2000ue,Buchel:2000cn,Bobev:2010de} (PW).
\end{itemize}
In the former case, the anisotropic thermal equilibrium states are characterized by the temperature $T$,
the background magnetic field $B$ and the renormalization scale $\mu$; in the latter case,
we have additionally a hypermultiplet mass scale $m$. 

Before we present results, we characterize  more precisely the models studied.
\nxt $\cftd$:  $\caln=4$ SYM has a global $SU(4)$ $R$-symmetry. In this model magnetic
field is turned on for the diagonal $U(1)$ of the $R$-symmetry. This is the model
of \cite{Endrodi:2018ikq}, see also \cite{DHoker:2009mmn}. See section \ref{diag} for the technical details.
\nxt $\cftstu$: Holographic duals of $\caln=4$ SYM with $U(1)^3\subset SU(4)$ global symmetry
are known as STU-models \cite{Behrndt:1998jd,Cvetic:1999ne}. In this conformal theory
the background magnetic field is turned on for one of the $U(1)$'s. This model is a
consistent truncation of $\caln=8$ five-dimensional gauged supergravity with
two scalar fields dual to two dimension $\Delta=2$ operators. As we show in section
\ref{stu}, in the presence of the background magnetic field these operators will develop
thermal expectation values. 
\nxt $\ncft$: As we show in section \ref{ncft}, within consistent truncation of
$\caln=8$ five-dimensional gauged supergravity presented in \cite{Bobev:2010de},
it is possible to identify a holographic dual to $\caln=2^*$ gauge theory with
a single $U(1)$ global symmetry. In this model the background magnetic field is turned on
in this $U(1)$. The label $m\in (0,+\infty)$ denotes the
hypermultiplet mass of the $\caln=2^*$ gauge theory. 
\nxt $\cftpw$:  This conformal gauge theory is a limiting case of the nonconformal
$\ncft$ model:
\[
\cftpw=\lim_{m/\sqrt{B}\to 0}\ \ncft\,.
\]
Its bulk gravitational dual contains two scalar fields dual to dimension $\Delta=2$ and
$\Delta=3$ operators of the $\caln=2^*$ gauge theory. As we show in section
\ref{m=0}, in the presence of the background magnetic field these operators will develop
thermal expectation values.   
\nxt $\cftminfty$:  This conformal gauge theory is a limiting case of the nonconformal
$\ncft$ model:
\[
\cftminfty=\lim_{m/\sqrt{B}\to \infty}\ \ncft\,.
\]
Its holographic dual can be obtained from the $\caln=8$ five dimensional gauged supergravity
of \cite{Bobev:2010de} using the "near horizon limit'' of
\cite{HoyosBadajoz:2010td}\footnote{See appendix D of \cite{Buchel:2019qcq} for details
of the isotropic (no magnetic field) thermal states of $\caln=2^*$ plasma in the limit
$m/T\to \infty$. The first hint that $\caln=2^*$ plasma in the infinite mass limit
is an effective five dimensional CFT appeared in \cite{Buchel:2007mf}.}, followed by the
uplift to six dimensions --- the resulting holographic dual is Romans $F(4)$ gauged supergravity
in six dimensions
\cite{Romans:1985tw,Cvetic:1999un}\footnote{See \cite{Chen:2019qib} for a recent discussion.}.
The six dimensional gravitational bulk contains a single scalar, dual to dimension
$\Delta=3$ operator of the effective CFT$ _5$. There is no conformal anomaly in odd dimensions.
Furthermore, there is no invariant dimension-five operator that can be constructed
only with the magnetic field strength --- as a result, the anisotropic stress-energy tensor of
$\cftminfty$ plasma is traceless, and is {\it free} from renormalization scheme ambiguities.
Details on the $\cftminfty$ model are presented in section \ref{minfty}.
The renormalization scheme-independence of $\cftminfty$ is a welcome feature: we will
use the pressure anisotropy \eqref{defr} of the theory as a benchmark to compare with the
other conformal and non-conformal models.

And now the results. There is no universal magnetoresponse.  Qualitatively,
among conformal/non-conformal models
we observe three different IR regimes (\ie when $T/\sqrt{B}$ is small):
\nxt In  $\cftd$ it is possible to reach deep IR, \ie  the  $T/\sqrt{B}\to 0$ limit. For
$T/\sqrt{B}\lesssim 0.1$ the thermodynamics is BTZ-like with the entropy
density\footnote{We independently reproduce this result.} 
\cite{DHoker:2009mmn}
\begin{equation}
s\to \frac{N^2}{3}\ B T\,,\qquad {\rm as}\qquad  \frac{T}{\sqrt{B}}\to 0\,.
\eqlabel{btz}
\end{equation}
\nxt Both in  $\cftpw$ and $\cftminfty$ (and in fact in all
$\ncft$ models) there is a terminal critical temperature $T_{crit}$ which
separates thermodynamically stable and unstable phases of the
anisotropic plasma. Remarkably, this $T_{crit}$ is universally determined by the
magnetic field $B$, (almost) independently\footnote{A very weak dependence
on the mass parameter has been also observed for the
equilibration rates in $\caln=2^*$ isotropic plasma in
\cite{Buchel:2015saa}.} of the mass parameter $m$ of $\ncft$:
\begin{center}
\begin{tabular}{ c c c c c c }
& $\cftpw$ & $\longrightarrow$ & $\ncft$ & $\longrightarrow$ & $\cftminfty$\\
\\
$\frac{T_{crit}}{\sqrt{B}}:$ & 0.29823(5) & $\longrightarrow$ & $[0.29823(6),0.30667(1)]$ & $\longrightarrow$ &0.30673(9)\\
\\
$\frac{m}{\sqrt{2B}}:$ & 0 & $\longrightarrow$ & $[1/100,10]$ & $\longrightarrow$ &$\infty$\,,\\
\end{tabular}
\end{center}
\ie the variation of ${T_{crit}}/{\sqrt{B}}$ with mass about its mean value is $2\%$ or less, see
Fig.~\ref{figure4} (left panel).
We leave the extensive study of this critical point to future work, and only point out that
the specific heat at constant $B$ at criticality has a critical exponent\footnote{The critical point with the same mean-field exponent $\alpha$ has been observed in isotropic thermodynamics 
of $\caln=2^*$ plasma with different masses for the bosonic and fermionic components
of the hypermultiplet \cite{Buchel:2010ys}.} $\alpha=\frac 12$:
\begin{equation}
c_B=-T \frac{\del^2 \calf}{(\del T)^2}\bigg|_{B}=\frac{\del s}{\del \ln T}\bigg|_{B}\
\propto\ \left(T-T_{crit}\right)^{-1/2}\,,
\eqlabel{alpha}
\end{equation}
where $\calf$ is the free energy density, see Fig.~\ref{alphaexp}. 
\nxt The $\cftstu$ model in the IR is different from the other ones.
We obtained reliable numerical results
in this model for $T/\sqrt{B}\gtrsim 0.06$:  we neither observe the critical point
as in the $\cftpw$ and $\cftminfty$ models, nor the BTZ-like behavior \eqref{btz} as in
the $\cftd$  model,
see Fig.~\ref{figure2} (left panel).

\begin{figure}[t]
\begin{center}
\psfrag{x}[cc][][0.7][0]{{${T}/{\sqrt{B}}$}}
\psfrag{y}[cc][][0.7][0]{{${P_T}/{P_L}$}}
\includegraphics[width=2.8in]{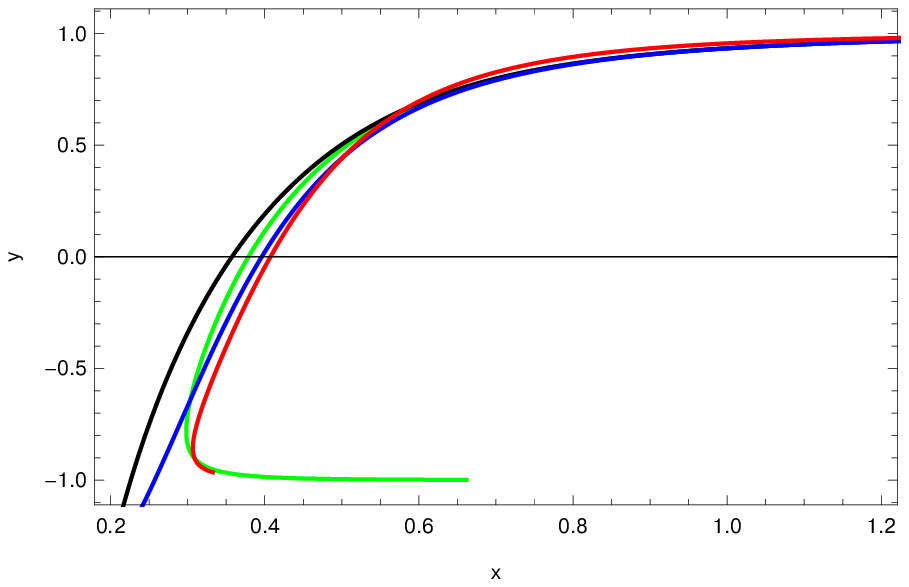}\qquad
\includegraphics[width=2.8in]{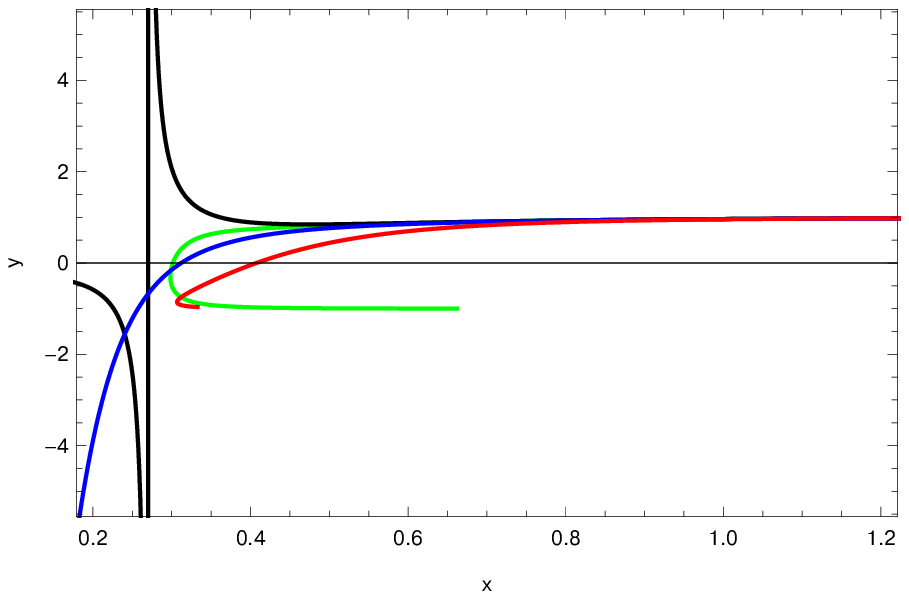}
\vskip 6pt
\includegraphics[width=2.8in]{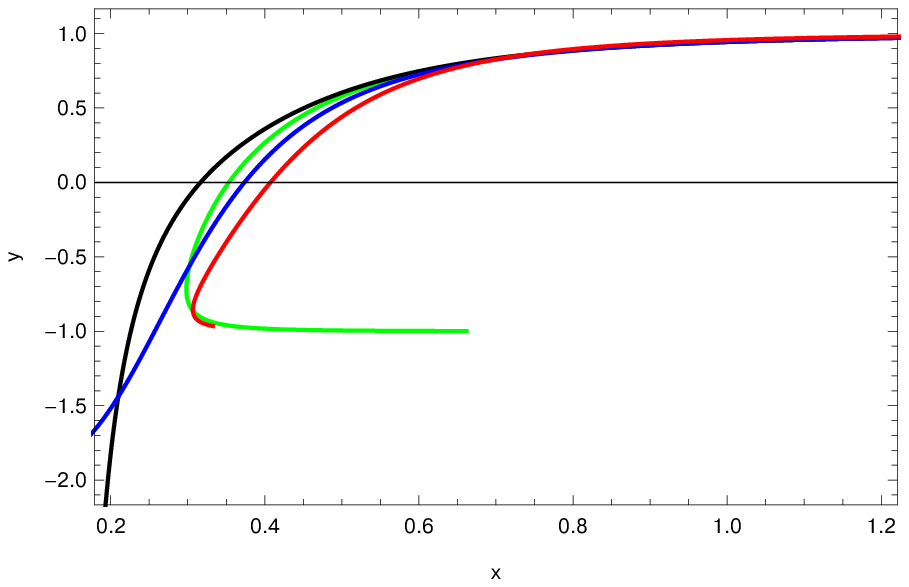}\qquad
\includegraphics[width=2.8in]{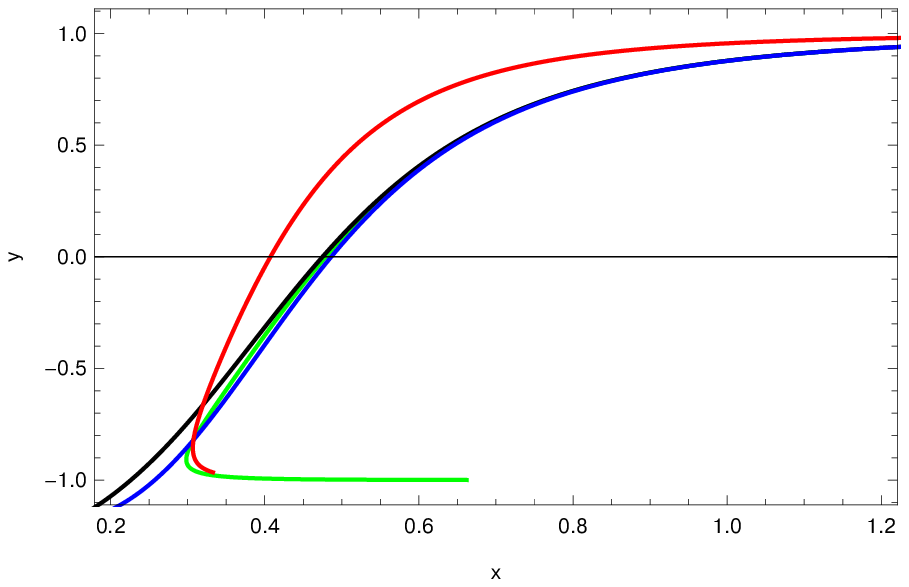}
\end{center}
  \caption{Anisotropy parameter $R=P_T/P_L$ for conformal models $\cftd$
  (black curves), $\cftstu$ (blue curves), $\cftpw$ (green curves) and $\cftminfty$
  (red curves) as a function of $T/\sqrt{B}$. $R_{\cftminfty}$ is renormalization scheme
  independent; for the other models there is a strong dependence on the renormalization
  scale $\delta=\ln \frac{B}{\mu^2}$: different panels represent different choices for $\delta$;
  all the models in the same panel have the same value of $\delta$, leading to identical
  high-temperature asymptotics, $T/\sqrt{B}\gg 1$. 
} \label{figure1}
\end{figure}

\begin{figure}[t]
\begin{center}
\psfrag{x}[cc][][1.0][0]{{${T}/{\sqrt{B}}$}}
\psfrag{y}[cc][][1.0][0]{{${P_T}/{P_L}$}}
\includegraphics[width=5.6in]{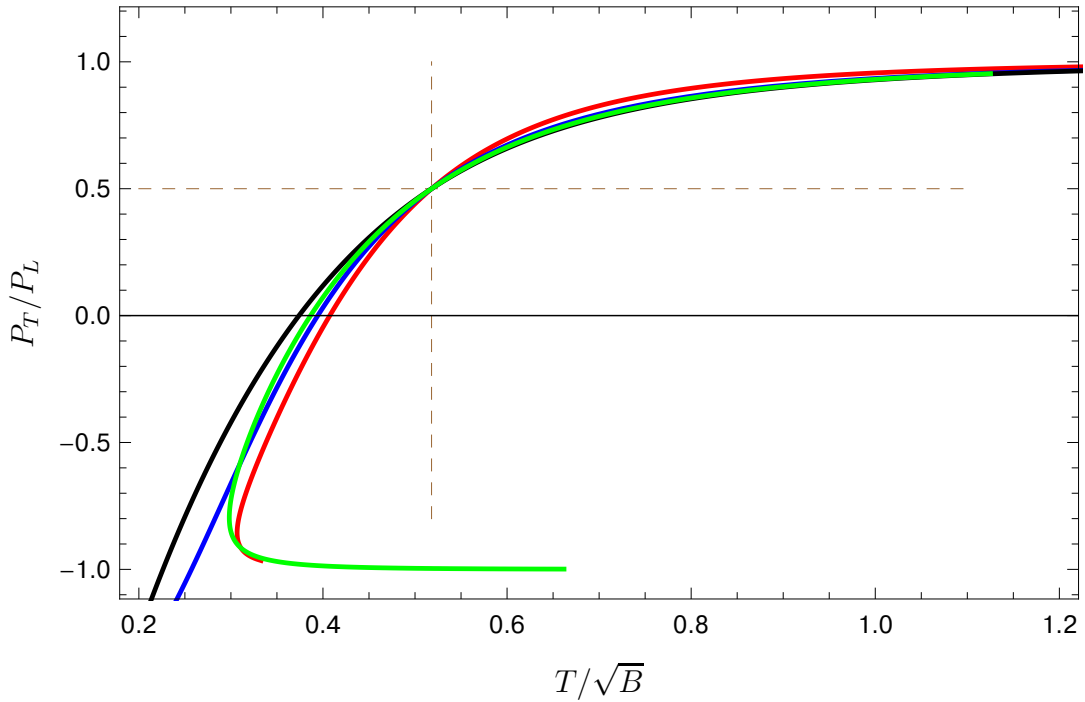}
\end{center}
  \caption{Renormalization scale $\delta$ is adjusted separately for
  the $\cftd$, $\cftstu$ and $\cftpw$ models (see \eqref{adjustcft})
  to ensure that in all these models the pressure anisotropy $R=0.5$
  occurs for the same value of $\frac{T}{\sqrt{B}}$ as in the $\cftminfty$ model
  (see \eqref{matchpoint}). This matching point is highlighted with the dashed
  brown lines. 
} \label{figure1a}
\end{figure}

In Fig.~\ref{figure1} we present the pressure anisotropy parameter $R$ \eqref{defr}
for the conformal theories: $\cftd$ (black curves), $\cftstu$ (blue curves), $\cftpw$ (green curves)
and $\cftminfty$ (red curves) as a function of\footnote{We use the same normalization
of the magnetic field in holographic models as in \cite{Endrodi:2018ikq}.} $T/\sqrt{B}$.
$R$ is renormalization scheme independent in the $\cftminfty$ model, while in the former three
conformal models it is  sensitive to
\begin{equation}
\delta\equiv \ln \frac{B}{\mu^2}\,,
\eqlabel{defdelta}
\end{equation}
where $\mu$ is the renormalization scale.  We performed high-temperature perturbative analysis, \ie as  $T/\sqrt{B}\gg 1$,
to ensure that the definition of $\delta$ is consistent across all the conformal models sensitive
to it, see appendix \ref{highT}. In the $\{$ top left,  top right, bottom left,  bottom right $\}$
panel of Fig.~\ref{figure1} we set  $\{ \delta=4\,, \delta=2.5\,, \delta=3.5\,, \delta=7\}$ (correspondingly)
for $R_\cftd$, $R_\cftstu$ and $R_\cftpw$ --- notice that while all the curves exhibit the same high-temperature
asymptotics, the anisotropy parameter $R$ is quite sensitive to $\delta$; in fact, $R_\cftd$
diverges for $\delta=2.5$ (because $P_L$ crosses zero with $P_T$ remaining finite).   
Varying $\delta$, it is easy to achieve $R_{\cftd}$, $R_\cftstu$ and $R_{\cftpw}$ in the IR to be
``to the left'' of the scheme-independent (red) curve $R_\cftminfty$ (top panels and the bottom left panel);
or "to the right'' of the scheme-independent (red) curve $R_\cftminfty$ (the bottom right panel).

In Fig.~\ref{figure1} we kept $\delta$ the same for the conformal models $\cftd$, $\cftstu$
and $\cftpw$.
This is very reasonable given that one can match $\delta$ across all the models
by comparing the UV, \ie $T/\sqrt{B}\gg 1$ thermodynamics (see appendix \ref{highT}) ---
there are no other scales besides $T$ and $B$, and thus by dimensional
analysis\footnote{The asymptotic $AdS_5$ radius $L$ always scales out from the
final formulas.},
\begin{equation}
P_{T/L}=T^4\ {\hat P}_{T/L}\biggl(\frac{T}{\sqrt{B}},\frac{\mu}{\sqrt{B}}\biggr) \,.
\eqlabel{pcft}
\end{equation}
If we give up on maintaining the same renormalization scale for all the
conformal models, it is easy to 'collapse' all the curves for the pressure anisotropy,
see Fig.~\ref{figure1a}. We will not perform sophisticated fits as in  \cite{Endrodi:2018ikq},
and instead, adjusting $\delta$ independently for each model,
we require that in all models the pressure anisotropy $R=0.5$ is attained
at the same value of $T/\sqrt{B}$ (represented by the dashed brown lines):
\begin{equation}
\frac{T}{\sqrt{B}}\bigg|_{\cftd,\cftstu,\cftpw}\
=\ \frac{T}{\sqrt{B}}\bigg|_{\cftminfty}= 0.51796(7)\,.
\eqlabel{matchpoint}
\end{equation}
Specifically, we find that \eqref{matchpoint} is true, provided
\begin{equation}
\left\{\delta_\cftstu\,, \delta_\cftd\,, \delta_\cftpw\right\}\ =\ \{3.9592(4)\,,\, 4.2662(0)\,,\,
4.1659(8)\}\,.
\eqlabel{adjustcft}
\end{equation}
In a nutshell, this is what was done in  \cite{Endrodi:2018ikq} to claim a universal
magnetoresponse for $R\gtrsim 0.5$. Rather, we interpret the collapse in
Fig.~\ref{figure1a} as nothing but a fitting artifact, possible due to a strong dependence
of the anisotropy parameter $R$ on the renormalization scale.

\begin{figure}[t]
\begin{center}
\psfrag{x}[cc][][1][0]{{${T}/{\sqrt{B}}$}}
\psfrag{y}[cc][][1][0]{{$s/s_{UV}$}}
\includegraphics[width=2.8in]{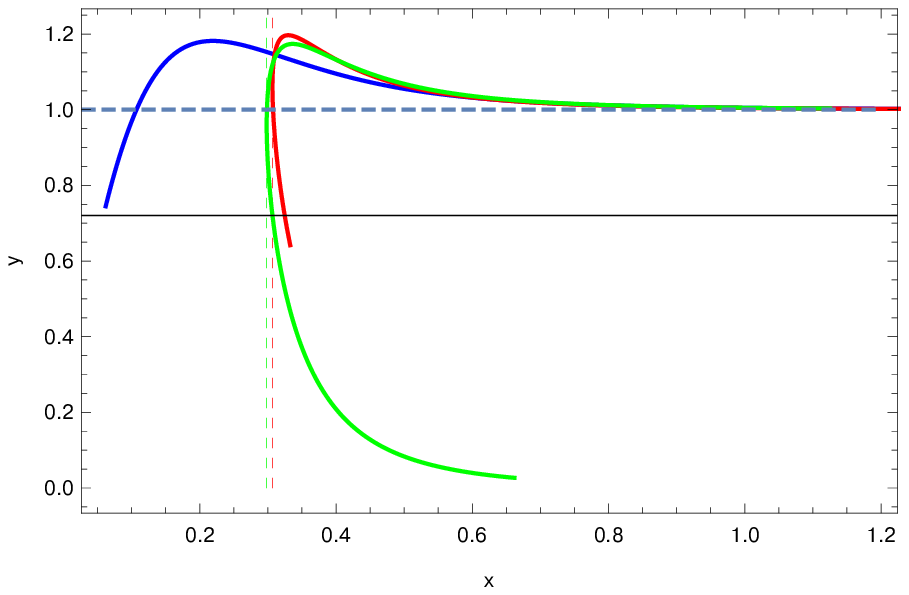}\qquad
\includegraphics[width=2.8in]{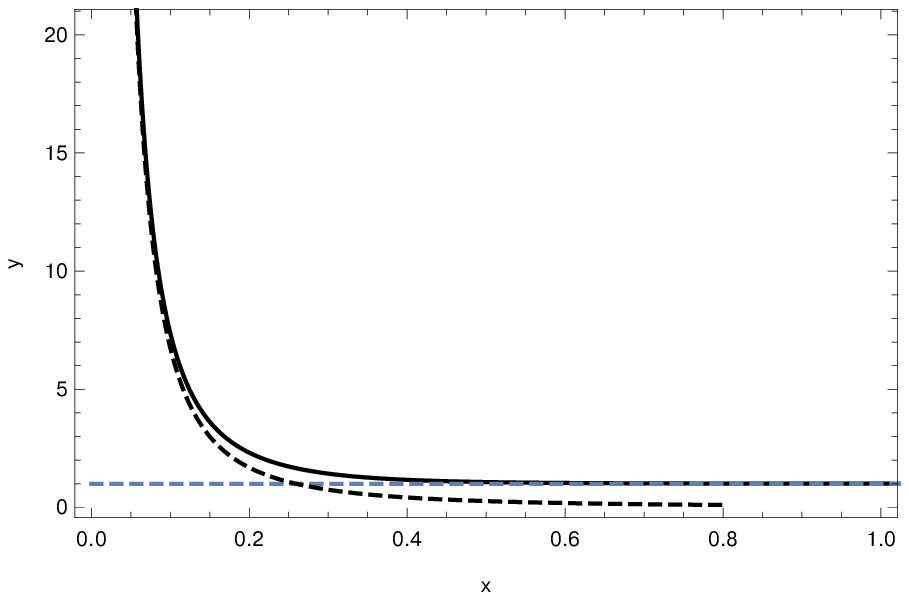}
\end{center}
  \caption{Entropy densities $s$ in conformal models, relative to the entropy densities
  of the UV fixed points $s_{UV}$ at the corresponding temperature (see \eqref{defsuv}),
  as functions of $T/\sqrt{B}$: $\cftd$ (black), $\cftstu$ (blue), $\cftpw$ (green)
  and $\cftminfty$ (red). Left panel: vertical dashed lines indicate critical temperatures
  $T_{crit}$ separating thermodynamically stable and unstable phases of
  $\cftpw$ (green) and $\cftminfty$ (red) models. Right panel: the dashed black
  line is the small-$T$ asymptote of the relative entropy in the  $\cftd$ model,
  see \eqref{btzass}.
} \label{figure2}
\end{figure}

To further see that there is no universal physics, we can compare  
renormalization scheme-independent anisotropic thermodynamic quantities of the models:
the entropy densities, see Fig.~\ref{figure2}. The color coding is as before:
 $\cftd$ (black curves), $\cftstu$ (blue curves), $\cftpw$ (green curves)
and $\cftminfty$ (red curves). We plot the entropy densities relative
to the entropy density of the UV fixed point at the corresponding
temperature (see eq.~(D.13) for the $\cftminfty$ model in \cite{Buchel:2019qcq}):
\begin{equation}
s_{UV}\bigg|_{\cftd\,, \cftstu\,, \cftpw}=\frac 12 \pi^2 N^2 T^3\,,\qquad (m\times s_{UV})\bigg|_{\cftminfty}=
\frac{432}{625} \pi^3 N^2 T^4\,.
\eqlabel{defsuv}
\end{equation}
The dashed vertical lines in the left panel indicate the terminal (critical temperature)
$T_{crit}/\sqrt{B}$ for $\cftpw$ (green) and $\cftminfty$ (red) models
which separates thermodynamically stable (top) and unstable
(bottom) branches.  Notice that $s/s_{UV}$ diverges for the $\cftd$ model as $T/\sqrt{B}\to 0$ ---
this is reflection of the IR BTZ-like thermodynamics \eqref{btz}; the dashed black
line is the IR asymptote
\begin{equation}
\frac{s}{s_{UV}}\bigg|_{\cftd}\to  \frac{2}{3\pi^2}\ \frac{B}{T^2}\,,\qquad {\rm as}\qquad
\frac{T}{\sqrt{B}}\to 0\,.
\eqlabel{btzass}
\end{equation}

\begin{figure}[t]
\begin{center}
\psfrag{x}[cc][][0.7][0]{{${T}/{\sqrt{B}}$}}
\psfrag{y}[cc][][0.7][0]{{${P_T}/{P_L}$}}
\includegraphics[width=2.8in]{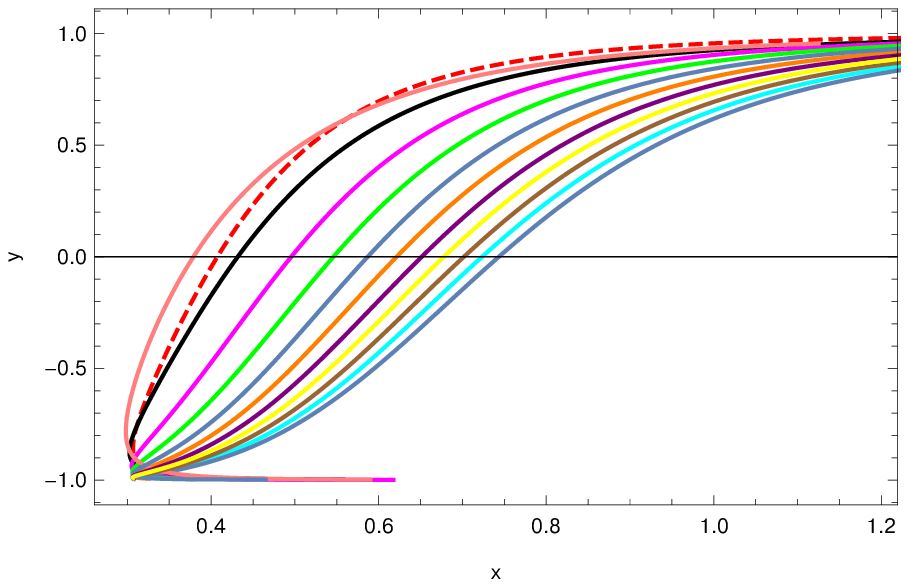}\qquad
\includegraphics[width=2.8in]{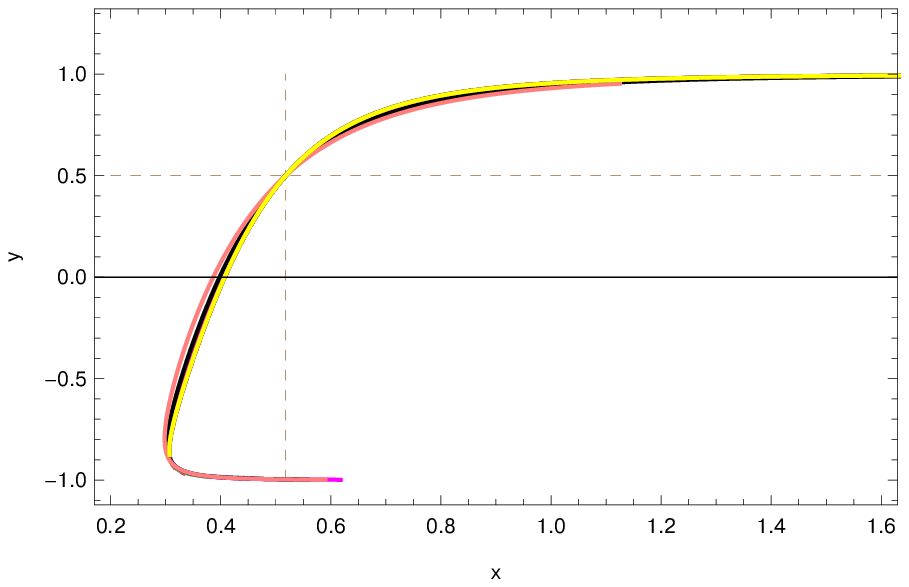}
\end{center}
  \caption{Anisotropy parameter $R=P_T/P_L$ for nonconformal models $\ncft$
  for select values of the hypermultiplet mass $m$, see \eqref{mchoice}, as a
  function of $T/\sqrt{B}$ (solid curves; from pink to dark blue as $m$ increases).
  The dashed red curve is a benchmark model $\cftminfty$ --- where the
  anisotropy parameter is renormalization scale independent. In the left panel the
  renormalization scale is set $\delta=4$ for all $\ncft$ models; in the right panel
  it is separately adjusted for each $\ncft$ model to ensure that all the curves
  pass through the matching point, highlighted with dashed brown lines.
} \label{figure30}
\end{figure}

In $\ncft$ models it is equally easy to 'collapse' the data for the pressure anisotropy.
In these models we have an additional scale $m$ --- the mass of the $\caln=2$
hypermultiplet. In the absence of the magnetic field, \ie for isotropic $\caln=2^*$
plasma, the thermodynamics is renormalization scheme-independent\footnote{Scheme-dependence
arises once we split the masses of the fermionic and bosonic components of the
$\caln=2^*$ hypermultiplet \cite{Buchel:2007vy}.} \cite{Buchel:2007vy}.
Once we turn on the magnetic field, there is a scheme-dependence. In Fig.~\ref{figure30}
we show the pressure anisotropy for $\caln=2^*$ gauge theory for select values of $m$
(solid curves from pink to dark blue),
\begin{equation}
\frac{m}{\sqrt{2B}}=\left\{\frac{1}{100},1,2,3,4,5,6,7,8,9,10\right\}\,.
\eqlabel{mchoice}
\end{equation}
The dashed red curve represents the anisotropy parameter of the conformal $\cftminfty$ model,
which is renormalization scheme-independent. In the left panel
the renormalization scale $\delta=4$ for all the $\ncft$ models. In the right panel,
we adjusted $\delta=\delta_m$ for each $\ncft$ model independently, so that the
pressure anisotropy $R_\ncft=0.5$ at the same temperature as in the $\cftminfty$ model,
see \eqref{matchpoint}. This matching point is denoted by dashed brown lines.

\begin{figure}[t]
\begin{center}
\psfrag{x}[cc][][0.7][0]{{${T}/{\sqrt{B}}$}}
\psfrag{y}[cc][][0.7][0]{{$s/s_{UV}$}}
\includegraphics[width=2.8in]{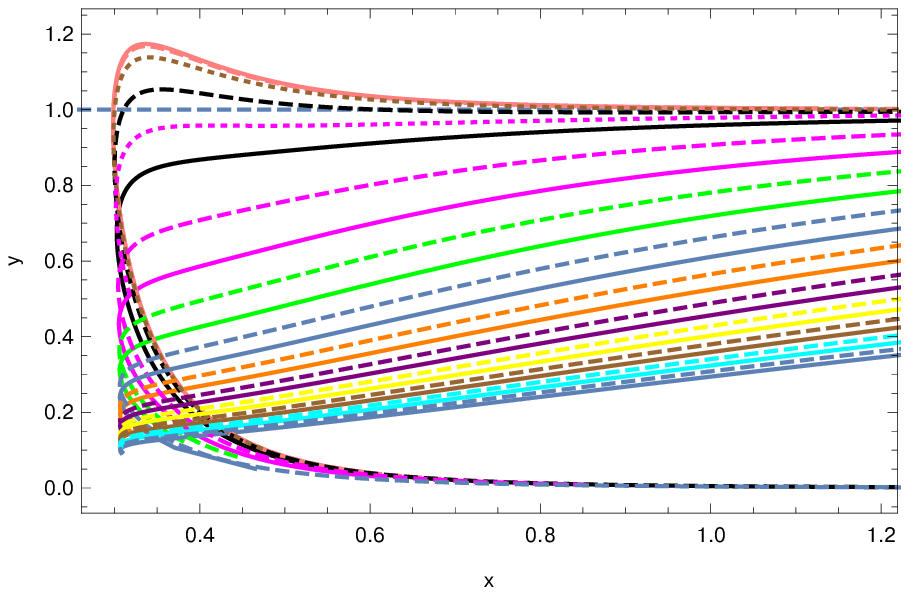}\qquad
\includegraphics[width=2.8in]{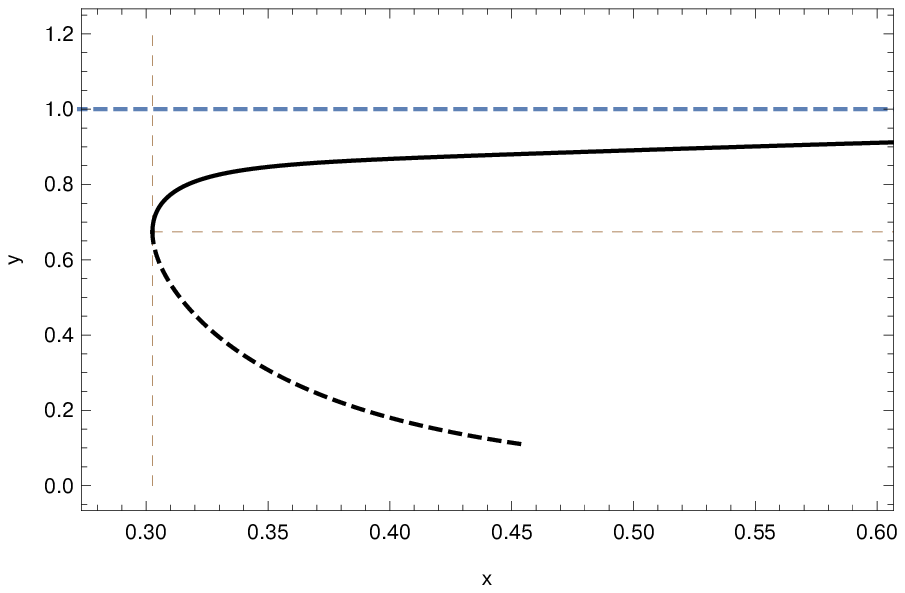}
\end{center}
  \caption{Left panel: entropy densities $s$ in $\ncft$  models, relative to the entropy density of
  the UV fixed point (the $\caln=4$ SYM in this case) $s_{UV}$ at
  the corresponding temperature (see \eqref{defsuv}), as functions of $T/\sqrt{B}$. Color coding
  of the solid curves agrees with that in Fig.~\ref{figure30} --- see \eqref{mchoice}
  for the set of the hypermultiplet masses. Additional dashed and dotted curves correspond to
  additional values of $m$, within the same interval \eqref{mchoice}. Each  $\ncft$ model
  has a terminal critical point. In the right panel we show this for the model with $m/\sqrt{2B}=1$:
  the brown lines identify the critical temperature $T_{crit}/\sqrt{B}$ and the relative entropy
  at the criticality $s^{crit}/s_{UV}$ (these quantities are presented in Fig.~\ref{figure4}).
 ``Top'' solid black curve denotes the thermodynamically stable branch and "bottom'' dashed black
 curve denotes the thermodynamically unstable branch (see Fig.~\ref{alphaexp} for further details).
} \label{figure3}
\end{figure}

\begin{figure}[t]
\begin{center}
\psfrag{x}[cc][][0.7][0]{{${T}/{\sqrt{B}}$}}
\psfrag{y}[cc][][0.7][0]{{$c_B/s$}}
\psfrag{z}[cc][][0.7][0]{{$(c_B/s)^{-2}$}}
\includegraphics[width=2.8in]{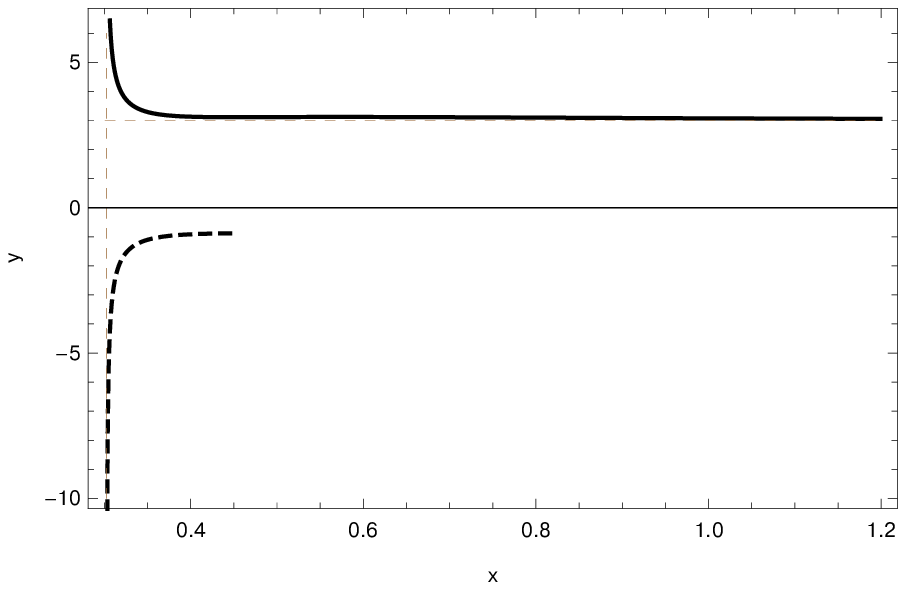}\qquad 
\includegraphics[width=2.8in]{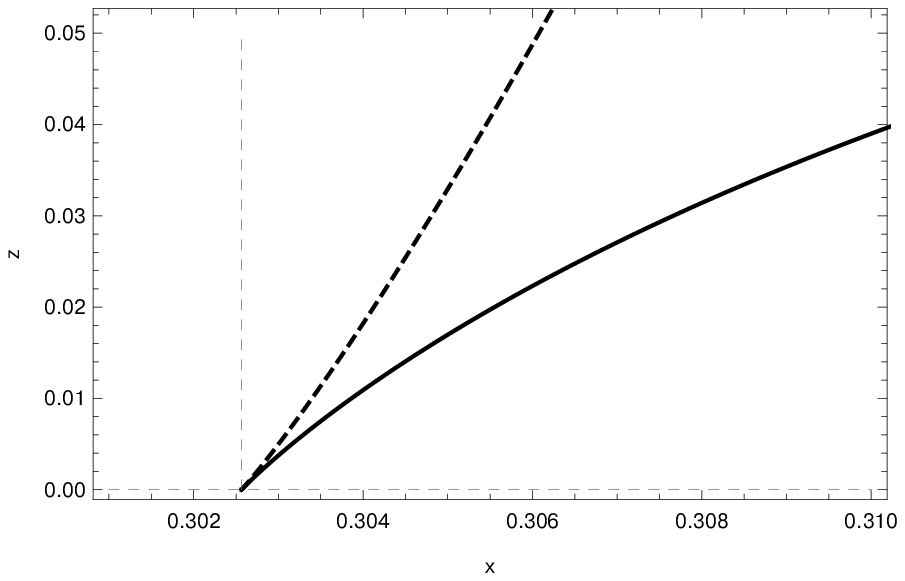}
\end{center}
  \caption{$\ncft$ model with $m/\sqrt{2B}=1$ is used to highlight phases of the
  anisotropic plasma. Following \eqref{alpha} we evaluate the constant-$B$ specific heat of the
  plasma. The dashed brown lines highlight the location of the critical point. Left panel:
  the specific heat diverges as one approaches the critical temperature; it is negative for the
  branch denoted by the dashed black curve (see also the right panel of Fig.~\ref{figure3}),
  indicating the thermodynamic instability. Right panel: $(c_B/s)^{-2}$ vanishes at
  criticality, with nonvanishing slope. This implies that the critical exponent $\alpha=\frac 12$,
  see \eqref{defalpha}. 
  } \label{alphaexp}
\end{figure}

As in conformal models, the entropy densities (which are renormalization scheme independent
thermodynamic quantities) are rather distinct, see left panel of Fig.~\ref{figure3}.
The color coding is as in Fig.~\ref{figure30}, except that we collected more
data\footnote{To have a better characterization of the critical points.} in addition to
\eqref{mchoice}: these are the dashed and dotted curves. The entropy density of the UV fixed point is
defined as in \eqref{defsuv}. All the $\ncft$ models studied, as well as the $\cftpw$ and
$\cftminfty$
conformal models, have a terminal critical point $T_{crit}$ that separates the
thermodynamically stable (top solid) and unstable (bottom dashed) branches, which we presented
for the  $\frac{m}{\sqrt{2B}}=1$ $\ncft$ model in the right panel. The dashed brown lines
identify the critical temperature $T_{crit}$ and the entropy density $s^{crit}$ at criticality.
In Fig.~\ref{alphaexp} we present results for the specific heat $c_B$ in this model defined
as in \eqref{alpha}. Indeed, the (lower) thermodynamically unstable branch has a negative specific
heat (left panel); approaching the critical temperature from above we observe
the divergence in the specific heat, both for the stable and the unstable branches.
To extract a critical exponent $\alpha$, defined as
\begin{equation}
c_B\ \propto\ \left(\frac{T}{T_{crit}}-1\right)^{-\alpha}\,,\qquad T\to T_{crit}+0\,,
\eqlabel{defalpha}
\end{equation}
we plot (right panel) the dimensionless quantity $c_B^2/s^2$ as a function of $T/\sqrt{B}$.
Both the stable
(solid) and the unstable (dashed) curves approach zero, signaling the divergence of the
specific heat at the critical temperature (vertical dashed brown line), with a finite slope ---
this implies that the critical exponent is
\begin{equation}
\alpha=\frac 12\,.
\eqlabel{resalpha}
\end{equation}

\begin{figure}[t]
\begin{center}
\psfrag{x}[cc][][0.7][0]{{${m}/{\sqrt{2B}}$}}
\psfrag{y}[cc][][0.7][0]{{$s^{crit}/s_{UV}$}}
\psfrag{z}[cc][][0.7][0]{{$T_{crit}/\sqrt{B}$}}
\includegraphics[width=2.8in]{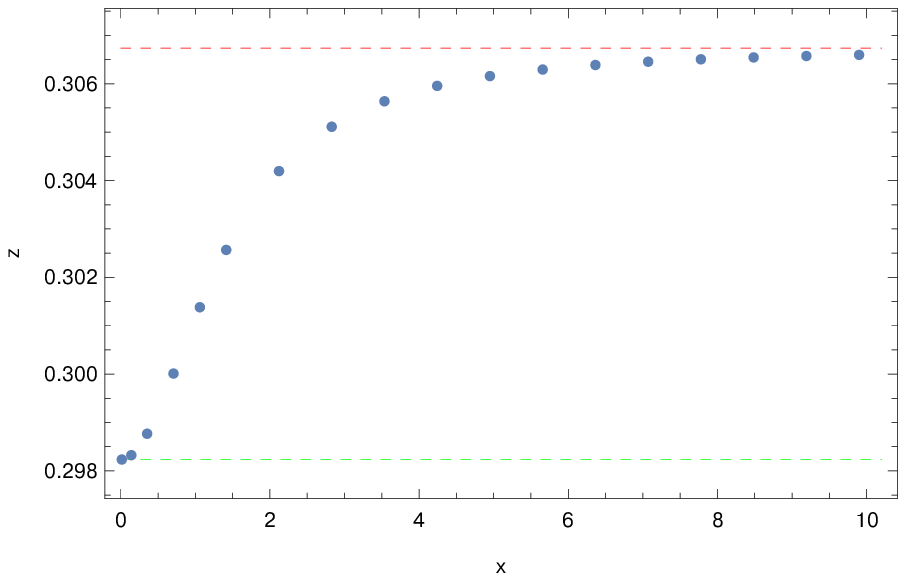}\qquad 
\includegraphics[width=2.8in]{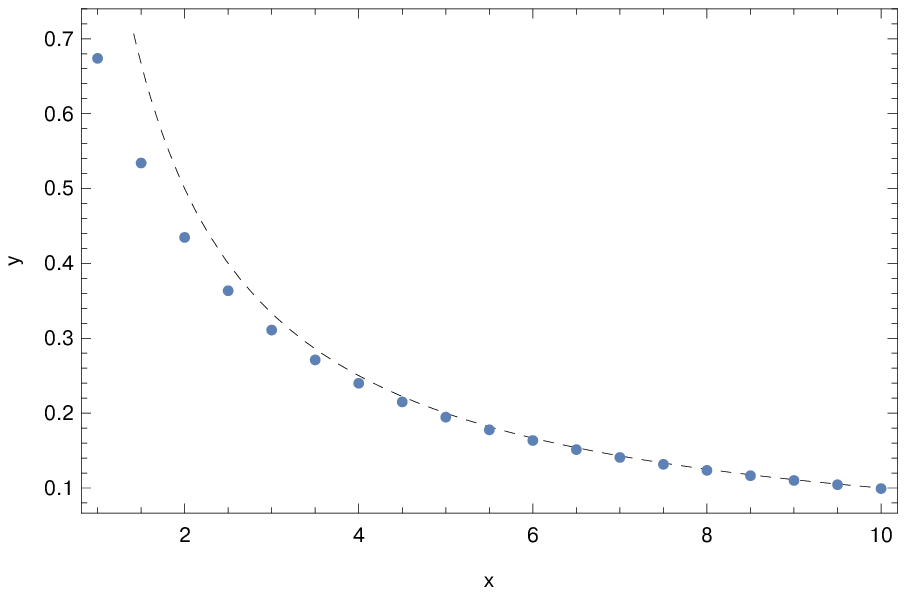}
\end{center}
  \caption{$\ncft$ models as well as the conformal models $\cftpw$ and $\cftminfty$
  have terminal critical temperature, separating thermodynamically stable and unstable phases.
  In the left panel we present $T_{crit}/\sqrt{B}$ as function of ${m}/\sqrt{2B}$;
  in the right panel we present the relative entropy at criticality $\gamma=s^{crit}/s_{UV}$
  \eqref{defgamma}. 
  The dots represent results for the $\ncft$ models;
the dashed horizontal lines (left panel) represent the critical temperature for the $\cftpw$ model
(green) and the $\cftminfty$ model (red). The dashed black curve (right panel) represents
the asymptote of $\gamma$ as $m/\sqrt{B}\to \infty$, see \eqref{gammainfty}. 
} \label{figure4}
\end{figure}

There is a remarkable universality of the critical points in $\ncft$ and
conformal $\cftpw$ and $\cftminfty$ models. In Fig.~\ref{figure4} (left panel) we present
the results for the critical temperature as a function of $m/\sqrt{2B}$ in $\ncft$ models
(points). The horizontal dashed lines indicate the location of the critical points
for the $\cftpw$ (green) and $\cftminfty$ (red) conformal models.
In the right panel the dots represent the relative entropy,
\begin{equation}
\gamma=\gamma(m/\sqrt{B})\equiv \frac{s^{crit}}{s_{UV}}\,,
\eqlabel{defgamma}
\end{equation}
at criticality for the $\ncft$ models.
Effectively, $\gamma$ as in \eqref{defgamma} measures the number of DOF at critical point in
anisotropic plasma relative to the number of DOF (or the central charge) of the UV fixed
point ($\caln=4$ SYM). 
The dashed black line is a simple asymptotic for $\gamma$ as $m/\sqrt{B}\to \infty$, $\gamma_\infty$,
\begin{equation}
\gamma_\infty=\frac{\sqrt{2B}}{m} \,.
\eqlabel{gammainfty}
\end{equation}
One can understand the origin of the asymptote \eqref{defgamma} from the fact that
$\ncft$ models in the large $m$ limit should resemble the conformal model $\cftminfty$;
thus, we expect that
$\gamma_\infty\approx \gamma_{\cftminfty}$.
Indeed,
\begin{equation}
\begin{split}
\gamma_{\cftminfty}=&\frac{s^{crit}_\cftminfty}{s_{UV,\cftd}}
=\underbrace{\frac{s^{crit}}{s_{UV}}\bigg|_{\cftminfty}}_{1.0603(7)}\ \times\qquad
\underbrace{\frac{s_{UV,\cftminfty}}{s_{UV,\cftd}}}_{\frac{864\pi}{625}\times \frac{T_{crit}}{m}}\\
=&1.0603(7)\ \times\  \frac{864\pi}{625\sqrt{2}}\ \times\
\underbrace{\frac{T_{crit}}{\sqrt{B}}}_{0.30673(9)}\ \times\ 
\frac{\sqrt{2B}}{m}
=0.99883(9)\ \times\  \frac{\sqrt{2B}}{m}\,,
\end{split}
\eqlabel{getgammainfty}
\end{equation}
where we extracted numerically the value of $\frac{s^{crit}}{s_{UV}}$ for the $\cftminfty$
conformal model, used \eqref{defsuv} to analytically compute the second factor in the first line, and 
substituted the numerical value for $T_{crit}/\sqrt{B}$ of the $\cftminfty$ model in the second line.

We now outline the rest of the paper, containing technical details necessary to obtain the results
reported above. In section \ref{model} we introduce the holographic theory of  \cite{Bobev:2010de}
and explain how the various models discussed here arise as  consistent truncations
of the latter: $\cftd$ in section \ref{diag}, $\cftstu$ in section \ref{stu}, and $\ncft$ in section
\ref{ncft}.  The conformal models $\cftpw$ and $\cftminfty$ are special limits of the $\ncft$ model
and are discussed in sections \ref{m=0} and \ref{minfty} correspondingly. Holographic renormalization
is by now a standard technique \cite{Skenderis:2002wp}, and we only present the
results for the boundary gauge theory observables. Our work is heavily numerical.
It is thus important to validate the numerical results in the limits where perturbative computations
(analytical or numerical) are available. We have performed such validations in appendix \ref{highT},
\ie when $\frac{T}{\sqrt{B}}\gg 1$.
We did not want to overburden the reader with details, and so we did not present the checks of the
agreement of the numerical parameters (\eg as in \eqref{stucond}) with the corresponding perturbative counterparts ---
but we have performed such checks in all models. There are further important constraints on the numerically
obtained energy density, pressure, entropy, etc.,  of the anisotropic plasma: the first law of the thermodynamics
$d\cale=T ds$ (at constant magnetic field and the mass parameter, if available), and the thermodynamic
relation between the free energy density and the longitudinal pressure $\calf=-P_L$. The latter relation
can be proved (see appendix \ref{proof}) at the level of the equations of motion, borrowing the holographic arguments
of \cite{Buchel:2003tz} used to establish the universality of the shear viscosity to the
entropy density in the holographic plasma models. Still, as the first law of thermodynamics,
it provides an important consistency check on the numerical data --- we verified these constraints
in all the models, both perturbatively in the high-temperature limit, to $\calo\left(\frac{B^4}{T^8}\right)$ inclusive,
see appendix \ref{highT}, and for finite values of $B/\sqrt{T}$, see appendix \ref{errorncft} -- once again,
we present only partial results of the full checks.

Our paper is a step in broadening the class of strongly coupled magnetized gauge theory plasmas
(both conformal and  massive) amenable to controlled holographic analysis. We focused on the
equation of state, extending the work of \cite{Endrodi:2018ikq}. The next step is to analyze the
magneto-transport in these models, in particular the magneto-transport at criticality.

\section{Technical details}\label{model}

The starting point for the holographic analysis is the effective action of \cite{Bobev:2010de}:
\begin{equation}
\begin{split}
&S_5=\frac{1}{4\pi G_5}\int_{\calm_5}d^5\xi\sqrt{-g} \biggl[
\frac R4-\frac 14 \biggl(\rho^4 \nu^{-4} F_{\mu\nu}^{(1)}F^{(1)\mu\nu}+\rho^4\nu^4F_{\mu\nu}^{(2)}
F^{(2)\mu\nu}+\rho^{-8}F_{\mu\nu}^{(3)}F^{(3)\mu\nu}\biggr)\\
&-\frac 12 \sum_{j=1}^4 \left(\del_\mu\phi_j\right)^2-3\left(\del_\mu\alpha\right)^2
-\left(\del_\mu\beta\right)^2-\frac 18 \sinh^2(2\phi_1)\left(
\del_\mu\theta_1+\left(A_\mu^{(1)}+A_\mu^{(2)}-A_\mu^{(3)}\right)
\right)^2\\
&-\frac 18 \sinh^2(2\phi_2)\left(
\del_\mu\theta_2+\left(A_\mu^{(1)}-A_\mu^{(2)}+A_\mu^{(3)}\right)\right)^2
-\frac 18 \sinh^2(2\phi_3)(
\del_\mu\theta_3+(-A_\mu^{(1)}+A_\mu^{(2)}\\
&+A_\mu^{(3)}))^2-\frac 18 \sinh^2(2\phi_4)\left(
\del_\mu\theta_4-\left(A_\mu^{(1)}+A_\mu^{(2)}+A_\mu^{(3)}\right)\right)^2
-\calp
\biggr]\,,
\end{split}
\eqlabel{model1}
\end{equation}
where the $F^{(J)}$ are the field strengths of the $U(1)$ gauge fields, $A^{(J)}$,
and $\calp$ is the scalar potential. We introduced
\begin{equation}
\rho\equiv e^\alpha\,,\qquad \nu\equiv e^\beta\,.
\eqlabel{model2}
\end{equation}
The scalar potential, $\calp$, is given in terms of a superpotential
\begin{equation}
\calp=\frac{g^2}{8}\biggl[\
\sum_{j=1}^4 \left(\frac{\del W}{\del\phi_j}\right)^2+\frac 16
\left(\frac{\del W}{\del\alpha}\right)^2
+\frac 12\left(\frac{\del W}{\del \beta}\right)^2\
\biggr]-\frac{g^2}{3} W^2\,,
\eqlabel{model3}
\end{equation}
where
\begin{equation}
\begin{split}
W=&-\frac{1}{4\rho^2\nu^2}\biggl[
\left(1+\nu^4-\nu^2\rho^6\right)\cosh(2\phi_1)+\left(-1+\nu^4+\nu^2\rho^6\right)
\cosh(2\phi_2)\\
&+\left(1-\nu^4+\nu^2\rho^6\right)\cosh(2\phi_3)+
\left(1+\nu^4+\nu^2\rho^6\right)\cosh(2\phi_4)
\biggr]\,.
\end{split}
\eqlabel{model4}
\end{equation}
In what follows we set gauged supergravity coupling $g=1$, this corresponds to setting the
asymptotic $AdS_5$ radius to $L=2$. The five dimensional gravitational constant
$G_5$ is related to the rank of the supersymmetric $\caln=4$ $SU(N)$ UV fixed point as
\begin{equation}
G_5=\frac{4\pi}{N^2}\,.
\eqlabel{defg5}
\end{equation}

The models discussed below, \ie $\cftd$, $\cftstu$ and  $\ncft$, have holographic duals which are
consistent truncations of \eqref{model1}. It would be interesting to study the stability of these
truncations following \cite{Balasubramanian:2013esa}.

\subsection{$\cftd$}\label{diag}

The holographic dual to the $\cftd$ conformal model is a consistent truncation of \eqref{model1}
with
\begin{equation}
\alpha=\beta=\phi_j=\theta_j=0\,,\qquad  A_\mu^{(1)}=A_\mu^{(2)}=A_\mu^{(3)}={\frac{2}{\sqrt{3}}}\ A_\mu\,,
\eqlabel{truncationd}
\end{equation}
leading to
\begin{equation}
\begin{split}
&S_\cftd=\frac{1}{16\pi G_5}\int_{\calm_5}d^5\xi\sqrt{-g} \biggl[
R- 4 F_{\mu\nu}F^{\mu\nu}+3
\biggr]\,,
\end{split}
\eqlabel{actiond}
\end{equation}
where we used the normalization of the bulk $U(1)$ to be consistent with \cite{Endrodi:2018ikq}.

This model has been extensively studied in \cite{DHoker:2009mmn,Endrodi:2018ikq} and we do not
review it here.

\subsection{$\cftstu$}\label{stu}
The holographic dual to the $\cftstu$ is a special case of the STU model \cite{Behrndt:1998ns,Behrndt:1998jd,Cvetic:1999ne}, a consistent truncation of the effective action \eqref{model1} with
\begin{equation}
\begin{split}
& \theta_j=\phi_j=0\,,
\end{split}
\end{equation}
leading to
\begin{equation}
\begin{split}
S_{STU}=&\frac{1}{4\pi G_5}\int_{\calm_5}d^5\xi\sqrt{-g} \biggl[
\frac R4-\frac 14 \biggl(\rho^4 \nu^{-4} F_{\mu\nu}^{(1)}F^{(1)\mu\nu}+\rho^4\nu^4F_{\mu\nu}^{(2)}
F^{(2)\mu\nu}\\&+\rho^{-8}F_{\mu\nu}^{(3)}F^{(3)\mu\nu}\biggr)
-3\left(\del_\mu\alpha\right)^2
-\left(\del_\mu\beta\right)^2-\calp_{STU}
\biggr]\,,
\end{split}
\eqlabel{modelstu}
\end{equation}
and the scalar potential
\begin{equation}
\calp_{STU}=-\frac{1}{4}(\rho^2\nu^2+\rho^2\nu^{-2}+\rho^{-4})\,.
\eqlabel{pstu}
\end{equation}
We would like to keep a  single bulk gauge field, so we can set two of them to zero
and work with the remaining one. The symmetries of the action allow us to
choose whichever gauge field we want. To see this, notice that the
action \eqref{modelstu} is invariant under
$F^{(1)}_{\mu\nu}\rightarrow F^{(2)}_{\mu\nu}$ together with
$\nu\rightarrow \nu^{-1}$. Moreover, \eqref{modelstu} with
$F^{(1)}_{\mu\nu}\equiv {2}F_{\mu\nu}$ and
$F^{(2)}_{\mu\nu}=F^{(3)}_{\mu\nu}=0$ is the same as with
$F^{(3)}_{\mu\nu}\equiv {2}F_{\mu\nu}$ and
$F^{(1)}_{\mu\nu}=F^{(2)}_{\mu\nu}=0$ for the gauge fields and with
the scalar field redefinitions $\rho\rightarrow \nu^{1/2}\rho^{-1/2}$
and $\nu\rightarrow\nu^{1/2}\rho^{1/2}$.
Thus, we arrive to the holographic dual of $\cftstu$ as
\begin{equation}
\begin{split}
S_{\cftstu}=&\frac{1}{4\pi G_5}\int_{\calm_5}d^5\xi\sqrt{-g} \biggl[
\frac R4-\rho^4 \nu^{-4} F_{\mu\nu}F^{\mu\nu}
-3\left(\del_\mu\alpha\right)^2
-\left(\del_\mu\beta\right)^2-\calp_{STU}
\biggr]\,,
\end{split}
\eqlabel{actionstu}
\end{equation}
where once again we used the normalization of the remaining gauge field as in \cite{Endrodi:2018ikq}. 

Solutions to the gravitational theory \eqref{actionstu} representing magnetic black branes
dual to anisotropic magnetized $\cftstu$ plasma correspond to the following background
ansatz\footnote{Note that we fixed the radial coordinate $r$ with the choice of the metric warp factor in
front of $dz^2$.}:
\begin{equation}
ds_5^2=-c_1^2\ dt^2+ c_2^2\ \left(dx^2+dy^2\right)+\left(\frac{r}{2}\right)^2\ dz^2+c_4^2\ dr^2\,,\qquad F=B\ dx\wedge dy\,,
\eqlabel{backstu}
\end{equation}
where all the metric warp factors $c_i$ as well as the bulk scalars $\rho$ and $\nu$
are functions of the radial coordinate $r$,
\begin{equation}
r\ \in\ [r_0,+\infty)\,,
\eqlabel{rrange}
\end{equation}
where $r_0$ is a location of a regular Schwarzschild horizon, and $r\to +\infty$ is the asymptotic
$AdS_5$ boundary. Introducing a new radial coordinate
\begin{equation}
x\equiv \frac{r_0}{r}\,,\qquad x\ \in\ (0,1]\,,
\eqlabel{defx}
\end{equation}
and denoting
\begin{equation}
\begin{split}
&c_1=\frac r2\ \left(1-\frac{r_0^4}{r^4}\right)^{1/2}\ a_1\,,\qquad c_2=\frac r2\ \ a_2\,,\qquad c_4=\frac 2r\ \left(1-\frac{r_0^4}{r^4}\right)^{-1/2}\ a_4\\
&B=\frac 12 r_0^2\ b\,,
\end{split}
\eqlabel{redefstu}
\end{equation}
we obtain the following system of ODEs (in a radial coordinate $x$, $'=\frac{d}{dx}$):
\begin{equation}
\begin{split}
&0=a_1'+\frac{a_1}{\nu^4 \rho^4 a_2^3 x (3 a_2-2 a_2' x) (1-x^4)}
\biggl(
\nu^4 \rho^4 a_2^2 x a_2' \left((x^4-1) x a_2'-2 (x^4-3) a_2\right)
\\
&-2 \nu^2 \rho^2 a_2^4 x^2 (x^4-1) \left(3 \nu^2 (\rho')^2+\rho^2 (\nu')^2\right)
-256 \rho^8 a_4^2 x^4 b^2+2 \nu^2 a_2^4 \left(a_4^2 (\nu^4 \rho^6+\rho^6+\nu^2\right)\\
&-3 \nu^2 \rho^4)
\biggr)\,,
\end{split}
\eqlabel{stu1}
\end{equation}
\begin{equation}
\begin{split}
&0=a_4'+\frac{a_4}{3 \nu^4 \rho^4 a_2^4 x (3 a_2-2 a_2' x) (x^4-1)} \biggl(
9 \nu^4 \rho^4 a_2^3 x^2 (x^4-1) (a_2')^2+6 \nu^2 \rho^2 a_2^5 x^2 (x^4-1)\\ &\times
\left(3 \nu^2 (\rho')^2+\rho^2 (\nu')^2\right)
+256 a_4^2 \rho^8 x^4 \left(9 a_2-4 a_2' x\right) b^2
-4 \nu^2 a_2^4 x (2 a_4^2 (\nu^4 \rho^6+\rho^6+\nu^2)\\
&+3 \nu^2 \rho^4 (x^4-2)) a_2'+6 \nu^2 a_2^5 (a_4^2 (\nu^4 \rho^6+\rho^6+\nu^2)-3 \nu^2 \rho^4)
\biggr)\,,
\end{split}
\eqlabel{stu2}
\end{equation}
\begin{equation}
\begin{split}
&0=a_2''-\frac{(a_2')^2}{a_2}-\frac{512 a_4^2 \rho^4 x^2 (3 a_2-a_2' x)}{
3\nu^4 a_2^4 (x^4-1)} b^2+\frac{a_2'}{3 x \rho^4 \nu^2 (x^4-1)} \biggl(
4 a_4^2 (\rho^6 \nu^4+\rho^6+\nu^2)\\
&+3 \rho^4 \nu^2 (x^4-1)\biggr)\,,
\end{split}
\eqlabel{stu3}
\end{equation}
\begin{equation}
\begin{split}
&0=\rho''-\frac{(\rho')^2}{\rho}+\frac{256a_4^2 \rho^4 x^2 (2 \rho' x+\rho)}
{3\nu^4 a_2^4 (x^4-1)} b^2+\frac{\rho'}{3 x \nu^2 \rho^4 (x^4-1)} \biggl(
4 a_4^2 (\rho^6 \nu^4+\rho^6+\nu^2)
\\&+3 \rho^4 \nu^2 (x^4-1)\biggr)
-\frac{a_4^2 (\rho^6 \nu^4+\rho^6-2 \nu^2)}{3\rho^3 \nu^2 x^2 (x^4-1)}\,,
\end{split}
\eqlabel{stu4}
\end{equation}
\begin{equation}
\begin{split}
&0=\nu''-\frac{(\nu')^2}{\nu}-\frac{256a_4^2 \rho^4 x^2 (3 \nu-2 \nu' x)}
{3\nu^4 a_2^4 (x^4-1)} b^2+\frac{\nu'}{3 \rho^4 \nu^2 x (x^4-1)} \biggl(
4 a_4^2 (\rho^6 \nu^4+\rho^6+\nu^2)\\
&+3 \rho^4 \nu^2 (x^4-1)\biggr)-\frac{a_4^2 \rho^2 (\nu^4-1)}{\nu x^2 (x^4-1)}\,.
\end{split}
\eqlabel{stu5}
\end{equation}
Notice that $r_0$ is completely scaled out from all the equations of motion.
Eqs.~\eqref{stu1}-\eqref{stu5} have to be solved subject to the following asymptotics:
\nxt in the UV, \ie as $x\to 0_+$,
\begin{equation}
\begin{split}
&a_1=1+a_{1,2}\ x^4+\calo(x^8\ln x)\,,\qquad a_2=1+\left(a_{2,2}-32 b^2 \ln x\right)\ x^4+ \calo(x^6)\,,\\
&a_4=1+\left(-a_{1,2}+\frac{64}{3}b^2-\frac43 n_1^2-4 r_1^2-2 a_{2,2}+64 b^2\ln x\right)\ x^4
+\calo(x^6)\,,\\
&\rho=1+r_1\ x^2+\calo(x^4)\,,\qquad \nu=1+n_1\ x^2+\calo(x^4)\,;
\end{split}
\eqlabel{uvstu}
\end{equation}
\nxt in the IR, \ie as $y\equiv 1-x\to 0_+$,
\begin{equation}
\begin{split}
&a_1=a_{1,h,0}+\calo(y)\,, \qquad  a_2=a_{2,h,0}+\calo(y)\,, \qquad \rho=r_{h,0}+\calo(y)\,,\qquad \nu=n_{h,0}+\calo(y)\,,\\
&a_4=\frac{3 a_{2,h,0}^2 r_{h,0}^2 n_{h,0}^2}{
(3 a_{2,h,0}^4 n_{h,0}^6 r_{h,0}^6+3 a_{2,h,0}^4 n_{h,0}^2 r_{h,0}^6
+96 b^2 r_{h,0}^8+3 a_{2,h,0}^4 n_{h,0}^4)^{1/2}}+\calo(y)\,.
\end{split}
\eqlabel{irstu}
\end{equation}
In total, given $b$ --- roughly the ratio $\frac{\sqrt{B}}{T}$, the asymptotic expansions are specified by 8 parameters:
\begin{equation}
\{a_{1,2}\,,\ a_{2,2}\,,\ r_1\,,\ n_1\,,\ a_{1,h,0}\,,\ a_{2,h,0}\,,\ r_{h,0}\,,\ n_{h,0}\}\,,
\eqlabel{stucond}
\end{equation}
which is the correct number of parameters necessary to provide a solution to a system of three
second order and two first order equations, $3\times 2+2\times 1=8$. The parameters 
$n_1$ and $r_1$ correspond to the expectation value of two dimension $\Delta=2$
operators of the boundary $\cftstu$; the other two parameters, $a_{1,2}$ and $a_{2,2}$, determine
the expectation value of its stress-energy tensor. Using the standard holographic renormalization
we find:
\begin{equation}
\begin{split}
&\langle T_{tt}\rangle\equiv \cale=\frac{r_0^4}{512\pi G_5}
\left(3-6 a_{1,2}-128 b^2 \ln r_0+128 b^2 \ln2+4 a_{2,2}+64 b^2\ \kappa \right)\,,
\\
&\langle T_{xx}\rangle=\langle T_{yy}\rangle\equiv P_T=\frac{r_0^4}{512\pi G_5}
\left(3-6 a_{1,2}-128 b^2 \ln r_0+128 b^2 \ln2+4 a_{2,2}+64 b^2\ \kappa\right)\,,
\\
&\langle T_{zz}\rangle\equiv P_L=\frac{r_0^4}{512\pi G_5}
\left(3-6 a_{1,2}-128 b^2 \ln r_0+128 b^2 \ln2+4 a_{2,2}+64 b^2\ \kappa\right)\,,
\end{split}
\eqlabel{tmunustu}
\end{equation}
for the components of the boundary stress-energy tensor, and
\begin{equation}
s=\frac{r_0^3 a_{2,h,0}^2}{32 G_5}\,,\qquad
T=\frac{
\sqrt{3} \left[a_{2,h,0}^4 n_{h,0}^2 (n_{h,0}^4 r_{h,0}^6+r_{h,0}^6+n_{h,0}^2)+32 b^2 r_{h,0}^8\right]^{1/2} a_{1,h,0} r_0}{12\pi r_{h,0}^2 n_{h,0}^2 a_{2,h,0}^2}\,,
\eqlabel{ststu}
\end{equation}
for the entropy density and the temperature.
Note that, as in $\caln=4$ SYM \cite{Endrodi:2018ikq},
\begin{equation}
\langle T^\mu_{\ \mu}\rangle=-\frac{r_0^4 b^2}{4\pi G_5}=-\frac{N^2}{4\pi^2}\ B^2\,,
\eqlabel{tracestu}
\end{equation}
where we used \eqref{redefstu} and \eqref{defg5}.
The (holographic) free energy density is given by the
standard relation
\begin{equation}
\calf=\cale-Ts\,.
\eqlabel{fstu}
\end{equation}
The constant parameter $\kappa$ in \eqref{tmunustu} comes from the finite counterterm of the holographic renormalization; we find it convenient to relate it to the renormalization scale $\mu$ in \eqref{defdelta} as   
\begin{equation}
\kappa=2\ln(2\pi \mu)\,.
\eqlabel{kappastu}
\end{equation}
As shown in appendix \ref{bstu}, the renormalization scheme choice \eqref{kappastu}
implies that in the high-temperature limit $T^2\gg B$,
\begin{equation}
R_\cftstu=1-\frac{4 B^2}{\pi^4 T^4}\ \ln\frac{T}{\mu\sqrt{2}}+\calo\left(\frac{B^4}{T^8}\ \ln^2 \frac{T}{\mu}\right)\,.
\eqlabel{rpertstu}
\end{equation}

We can not solve the equations \eqref{stu1}-\eqref{stu5} analytically; adapting numerical techniques
developed in \cite{Aharony:2007vg}, we solve these equations (subject to the asymptotics
\eqref{uvstu} and \eqref{irstu}) numerically. The results of numerical analysis are data
files assembled of parameters \eqref{stucond}, labeled by $b$.  
It is important to validate the numerical data (in addition to the standard error analysis).
There are two important constraints that we verified for $\cftstu$ (and in fact all the other
models):
\begin{itemize}
\item The first law of thermodynamics (FL), $d\cale/(T ds)-1$ (with $B$ kept fixed), leads to the
differential constrain on data sets \eqref{stucond} (here $'=\frac{d}{db}$):
\begin{equation}
\begin{split}
{\rm FL:}\ 0=&\frac{
\sqrt3 r_{h,0}^2 n_{h,0}^2 a_{2,h,0}( (2 a_{2,2}'-3 a_{1,2}') b+32 b^2+6 a_{1,2}-4 a_{2,2}-3)}{(4 a_{2,h,0}' b-3 a_{2,h,0})a_{1,h,0}\sqrt{a_{2,h,0}^4 n_{h,0}^2((n_{h,0}^4+1) r_{h,0}^6
+n_{h,0}^2)+32 b^2 r_{h,0}^8} }-1\,.
\end{split}
\eqlabel{flstu}
\end{equation}
\item Anisotropy introduced by the external magnetic field results in $P_T\ne P_L$. From the
elementary anisotropic thermodynamics (see \cite{Endrodi:2018ikq} for a recent review), the free energy density of the
system $\calf$ is given by 
\begin{equation}
\calf=-P_L\qquad \Longrightarrow\qquad 0=\frac{\cale+P_L}{sT}-1\,.
\eqlabel{fstu2}
\end{equation}
We emphasize that holographic renormalization (even anisotropic one) naturally enforces
\eqref{fstu} (see \cite{Buchel:2004hw} for one of the first demonstrations),
but not \eqref{fstu2}. In appendix \ref{proof} we present a holographic
proof\footnote{The proof follows the same steps as in the first proof of the
universality of the shear viscosity to the entropy density in holography \cite{Buchel:2003tz}.} of the
thermodynamic relation (TR) \eqref{fstu2}. Applying it to $\cftstu$ model we arrive at the constraint
\begin{equation}
{\rm TR:}\ 0=\frac{\sqrt{3}(1-2 a_{1,2})  r_{h,0}^2 n_{h,0}^2}{
a_{1,h,0}\sqrt{a_{2,h,0}^4 n_{h,0}^2 ((n_{h,0}^4+1) r_{h,0}^6
+n_{h,0}^2)+32 b^2 r_{h,0}^8} }-1\,.
\eqlabel{fstu3}
\end{equation}
\end{itemize}

In appendix \ref{bstu} we have verified FT and TR in the $\cftstu$ model to order
$\calo(b^4) \sim \calo(B^4/T^8)$ inclusive\footnote{Additionally, as in the $\ncft$ model with
$m/\sqrt{2B}=1$ (see appendix \ref{errorncft}),  we checked
both relations for finite $b$.}.   

Technical details presented here are enough to generate the $\cftstu$ model plots reported in
section \ref{intro}.

\subsection{$\ncft$}\label{ncft}

There is a simple consistent truncation of the effective action \eqref{model1}
to that of the PW action \cite{Pilch:2000ue}, supplemented with a single bulk $U(1)$ gauge field.
Indeed, setting
\begin{equation}
\begin{split}
&\beta=0 \Longrightarrow \nu=1\,,\qquad \phi_2=\phi_3\equiv\chi\,,\qquad \phi_1=\phi_4=0\,,\\
&A^{(1)}=A^{(2)}\equiv \sqrt{2} A\,,\qquad A^{(3)}=0\,,\qquad \theta_J=0\,.
\end{split}
\end{equation}
we find
\begin{equation}
\begin{split}
{S}_\ncft=\frac{1}{4\pi G_5}\int_{\calm_5} d^5 \xi \sqrt{-g}\biggl[
\frac R4-3\left(\del_\mu\alpha\right)^2-\left(\del_\mu\chi\right)^2-\calp_{PW}
-{\rho^4} F_{\mu\nu}F^{\mu\nu}
\biggr]\,,
\end{split}
\eqlabel{pw1}
\end{equation}
where $\calp_{PW}$ is the Pilch-Warner scalar potential of the gauged supergravity:
\begin{equation}
\begin{split}
\calp_{PW}=&\frac{1}{48} \left(\frac{\del W_{PW}}{\del\alpha}\right)^2+\frac{1}{16}
\left(\frac{\del W_{PW}}{\del\chi}\right)^2-\frac 13 W_{PW}^2\,,\\
W_{PW}=&-\frac{1}{\rho^2}-\frac 12 \rho^4 \cosh(2\chi)\,.
\end{split}
\eqlabel{pw2}
\end{equation}
We use the same holographic background ansatz, the same radial coordinate $x$,
as for the $\cftstu$ model
\eqref{backstu}-\eqref{redefstu}; except that now we have the bulk scalar fields
$\alpha$ and $\chi$ (here $'=\frac{d}{dx}$):  
\begin{equation}
\begin{split}
&0=a_1'+\frac{2 a_1 a_2 x}{3 a_2-2 a_2' x} \left((\chi')^2+3 (\alpha')^2\right)
+\frac{a_1a_2'}{2a_2} -\frac{a_1 (x^4-9)}{4x (x^4-1)}+\frac{64 a_1 a_4^2 e^{4 \alpha} x^3b^2}
{a_2^3 (3 a_2-2 a_2' x) (x^4-1)}\\
&-\frac{a_1 a_2a_4^2}{8x (3a_2-2 a_2' x) (x^4-1)}
\biggl(2 e^{8 \alpha}+16 e^{-4 \alpha}-e^{8 \alpha-4 \chi}+16 e^{2 \alpha
+2 \chi}+16 e^{2 \alpha-2 \chi}-e^{8 \alpha+4 \chi}
\biggr)\\
&+\frac{3a_1 a_2}{4x (3a_2-2 a_2' x)}\,,
\end{split}
\eqlabel{pw1e}
\end{equation}
\begin{equation}
\begin{split}
&0=a_4'+\frac{2 a_4 a_2 x}{3 a_2-2 a_2' x} \left(
(\chi')^2+3 (\alpha')^2\right)
-\frac{3a_4a_2'}{2a_2}+\frac{64a_4^3 e^{4 \alpha} x^3 (9a_2-4 a_2' x)b^2}
{3a_2^4 (3 a_2-2 a_2' x) (x^4-1)}
\\
&+\frac{a_4^3(3a_2-4 x a_2')}{24x (3a_2-2 a_2'x) (x^4-1)} \biggl(
2 e^{8 \alpha}  +16 e^{2 \alpha+2 \chi}  
- e^{8 \alpha-4 \chi}  +16 e^{2 \alpha-2 \chi}  
+16  e^{-4 \alpha} -e^{8 \alpha+4 \chi} \biggr)\\
&-\frac{a_4 (12 a_2-a_2' x (x^4+7) x)}{2(x^4-1) (3a_2-2 a_2' x) x}\,,
\end{split}
\eqlabel{pw2e}
\end{equation}
\begin{equation}
\begin{split}
&0=a_2''-\frac{(a_2')^2}{a_2}-\frac{128a_4^2 e^{4 \alpha} x^2 (3a_2-a_2' x)b^2}
{3a_2^4 (x^4-1)} 
+\frac{a_2'}{12x (x^4-1)} \biggl(
12 (x^4-1)+ a_4^2  (2 e^{8 \alpha}\\
&+16  e^{2 \alpha+2 \chi}- e^{8 \alpha-4 \chi}+16
e^{2 \alpha-2 \chi}+16  e^{-4 \alpha}- e^{8 \alpha+4 \chi})
\biggr)\,,
\end{split}
\eqlabel{pw3e}
\end{equation}
\begin{equation}
\begin{split}
&0=\alpha''+\frac{64a_4^2 e^{4 \alpha} x^2 (2 \alpha' x+1)b^2}{3a_2^4 (x^4-1)}
+\frac{\alpha'}{12x (x^4-1)} \biggl(12 (x^4-1)+a_4^2 (2  e^{8 \alpha}+16
e^{2 \alpha+2 \chi}\\
&- e^{8 \alpha-4 \chi}+16  e^{2 \alpha-2 \chi}+16  e^{-4 \alpha}- e^{8 \alpha+4 \chi})
\biggr)
-\frac{a_4^2}{12x^2 (x^4-1)} \biggl(2 e^{8 \alpha}+4 e^{2 \alpha+2 \chi}-e^{8 \alpha-4 \chi}\\
&+4 e^{2 \alpha-2 \chi}
-8 e^{-4 \alpha}-e^{8 \alpha+4 \chi}\biggr)\,,
\end{split}
\eqlabel{pw4e}
\end{equation}
\begin{equation}
\begin{split}
&0=\chi''+\frac{128a_4^2 \chi' e^{4 \alpha} x^3 b^2}{3a_2^4 (x^4-1)}
+\frac{\chi'}{12x (x^4-1)} \biggl(
12 (x^4-1)+a_4^2
(2  e^{8 \alpha}+16  e^{2 \alpha+2 \chi}- e^{8 \alpha-4 \chi}
\\&+16  e^{2 \alpha-2 \chi}+16  e^{-4 \alpha}- e^{8 \alpha+4 \chi})
\biggr) -\frac{a_4^2\left(
8 e^{2 \alpha+2 \chi}+e^{8 \alpha-4 \chi}-8 e^{2 \alpha-2 \chi}-e^{8 \alpha+4 \chi}\right)}{8x^2 (x^4-1)} \,.
\end{split}
\eqlabel{pw5e}
\end{equation}
As in the $\cftstu$ model, $r_0$ is completely scaled out from all the equations of motion.
Eqs.~\eqref{pw1e}-\eqref{pw5e} have to be solved subject to the
following asymptotics:
\nxt in the UV, \ie as $x\to 0_+$,
\begin{equation}
\begin{split}
&a_1=1-x^4 \left(4 \alpha_{1,0}^2+2 \alpha_{1,0} \alpha_{1,1}+\frac{\alpha_{1,1}^2}{2}-\frac{64b^2}{3} 
+2 \chi_0 \chi_{1,0}+2 a_{2,2,0}+a_{4,2,0}\right)+\calo(x^6)\,,\\
&a_2=1+x^4 \left(-32 b^2 \ln x+a_{2,2,0}\right)+\calo\left(x^6 \ln x\right)\,,\\
&a_4=1-\frac23 x^2 \chi_0^2+x^4 \biggl(
-4 \alpha_{1,1}^2\ \ln^2 x +\left(-8 \alpha_{1,0} \alpha_{1,1}+64 b^2-\frac83 \chi_0^4-2 \alpha_{1,1}^2\right) \ln x
\\&+a_{4,2,0}\biggr)
+\calo\left(x^6 \ln^3 x\right)\,,
\\
&\alpha=x^2 \left(\alpha_{1,1}\ \ln x+\alpha_{1,0}\right)+\calo\left(x^4 \ln^2 x\right)\,,
\\&\chi=\chi_0 x+\left(
\frac43 \chi_0^3\ \ln x+\chi_{1,0}\right) x^3+\calo\left(x^5 \ln^2 x\right)\,;
\end{split}
\eqlabel{uvpw}
\end{equation}
\nxt in the IR, \ie as $y\equiv1-x\to 0_+$,
\begin{equation}
\begin{split}
&a_1=a_{1,h,0}+\calo(y)\,,\
a_2=a_{2,h,0}+\calo(y)\,,\ \alpha=\ln r_{h,0}+\calo(y)\,,\ \chi=\ln c_{h,0}+\calo(y)\,, \\
&a_4={4 \sqrt3 a_{2,h,0}^2 r_{h,0}^2 c_{h,0}^2}\ \biggl(
a_{2,h,0}^4 (r_{h,0}^6 (16 c_{h,0}^6-r_{h,0}^6 (1-c_{h,0}^4)^2)+16 c_{h,0}^2 (r_{h,0}^6+c_{h,0}^2))\\
&\qquad\qquad+512 b^2 c_{h,0}^4 r_{h,0}^8
\biggr)^{-1/2}+\calo(y)\,.
\end{split}
\eqlabel{irpw}
\end{equation}
The non-normalizable coefficients $\alpha_{1,1}$ (of the dimension $\Delta=2$ operator) 
and $\chi_0$ (of the dimension $\Delta=3$ operator) are related to the masses of the bosonic and
the fermionic components of the hypermultiplet of $\caln=2^*$ gauge theory. When both masses are the
same  (see \cite{Buchel:2007vy})
\begin{equation}
\alpha_{1,1}=\frac{2}{3}\ \chi_0^2\,.
\eqlabel{mbmf}
\end{equation}
Furthermore, carefully matching to the extremal PW solution  \cite{Pilch:2000ue,Buchel:2000cn}
(following the same procedure as in \cite{Buchel:2007vy}) we find
\begin{equation}
\frac{B}{m^2}=\frac{2b}{\chi_0^2}\,, 
\eqlabel{bm2}
\end{equation}
where $m$ is the hypermultiplet mass. We find it convenient to use 
\begin{equation}
\eta\equiv \frac{m}{\sqrt{2B}}\qquad \Longrightarrow \qquad \chi_0=2\sqrt{b}\ \eta \,,
\eqlabel{defeta}
\end{equation}
to label different mass parameters in $\ncft$ models, see \eqref{mchoice}. 
In total, given $\eta$ and $b$, the asymptotics expansions are specified by
8 parameters:
\begin{equation}
\{a_{2,2,0}\,,\, a_{4,2,0}\,,\, \alpha_{1,0}\,,\, \chi_{1,0}\,,\, a_{1,h,0}\,,\, a_{2,h_0}\,,\,
r_{h,0}\,,\, c_{h,0}\}\,,
\eqlabel{pwcond}
\end{equation}
which is the correct number of parameters necessary to provide a solution to a system of three
second order and two first order equations, $3\times 2+2\times 1=8$. Parameters 
$\alpha_{1,0}$ and $\chi_{1,0}$ correspond to the expectation values of dimensions $\Delta=2$ ($\calo_2$)
and $\Delta=3$ ($\calo_3$) operators (correspondingly) of the boundary $\ncft$;
the other two parameters, $a_{2,2,0}$ and $a_{4,2,0}$, determine
the expectation value of its stress-energy tensor. Using the standard holographic renormalization
\cite{Buchel:2012gw} we find:
\begin{equation}
\begin{split}
&\langle T_{tt}\rangle\equiv \cale=\frac{r_0^4}{1536\pi G_5} \biggl(9-64 b^2 \left(\eta^4+6 \ln r_0-6 \ln 2
-3 \kappa+6\right)+192 \alpha_{1,0} b \eta^2+72 \alpha_{1,0}^2\\
&\qquad\qquad\qquad\qquad\qquad+48 a_{2,2,0}+18 a_{4,2,0}+48 \sqrt{b} \eta \chi_{1,0}\biggr)\,,
\\
&\langle T_{xx}\rangle=\langle T_{yy}\rangle\equiv P_T=\frac{r_0^4}{4608\pi G_5} \biggl(9
-64 b^2 (-7 \eta^4+18 \ln r_0-18 \ln 2-9 \kappa+15)\\
&\qquad\qquad\qquad\qquad\qquad\qquad-192 \alpha_{1,0} b \eta^2
+72 \alpha_{1,0}^2+144 \sqrt{b} \eta \chi_{1,0}+72 a_{2,2,0}+18 a_{4,2,0}\biggr)\,,
\\
&\langle T_{zz}\rangle\equiv P_L=\frac{r_0^4}{4608\pi G_5} \biggl(9+64 b^2 \left(
7 \eta^4+18 \ln r_0-18 \ln 2-9 \kappa-6\right)-192 \alpha_{1,0} b \eta^2\\
&\qquad\qquad\qquad\qquad\qquad\qquad+72 \alpha_{1,0}^2
+144 \sqrt{b} \eta \chi_{1,0}+18 a_{4,2,0}\biggr)\,,
\end{split}
\eqlabel{tmununcft}
\end{equation}
for the components of the boundary stress-energy tensor,
\begin{equation}
\calo_2=\frac{r_0^2}{8\pi G_5}\ \left(\alpha_{1,0}-\frac 23\eta^2 b\right)\,,\qquad \calo_3=-\frac{r_0^3}{16\pi G_5}
\left(\chi_{1,0}+\frac{8}{3} \eta^3 b^{3/2}\right)\,,
\eqlabel{o2o3vevs}
\end{equation}
for the expectation values of the relevant operators, and
\begin{equation}
\begin{split}
&s=\frac{r_0^3 a_{2,h,0}^2}{32G_5}\,,\qquad  T=\frac{\sqrt{3} r_0 a_{1,h,0}}{48\pi a_{2,h,0}^2 c_{h,0}^2 r_{h,0}^2}
\biggl[a_{2,h,0}^4 (16 c_{h,0}^2 (r_{h,0}^6+c_{h,0}^2)
-r_{h,0}^6 ((c_{h,0}^4-1)^2 r_{h,0}^6\\
&-16 c_{h,0}^6))+512 b^2 c_{h,0}^4 r_{h,0}^8\biggr]^{1/2}\,,
\end{split}
\eqlabel{stncft}
\end{equation}
for the entropy density and the temperature.
Note that, as expected \cite{Buchel:2012gw},
\begin{equation}
\begin{split}
\langle T^\mu_{\ \mu}\rangle=&-\frac{r_0^4}{4 \pi G_5} \biggl(
b^2 \left(1-\frac43 \eta^4\right)+\alpha_{1,0} b \eta^2-\frac14 \sqrt{b} \eta \chi_{1,0}\biggr)\\
=&-2 m^2\ \calo_2-m\ \calo_3-\frac{N^2}{4\pi^2} B^2\,,
\end{split}
\eqlabel{tracencft}
\end{equation}
where in the second equality we used \eqref{redefstu}, \eqref{defg5}, \eqref{o2o3vevs} and \eqref{defeta}. The (holographic) free energy density is directly given by the standard relation \eqref{fstu}.
The constant parameter $\kappa$ in \eqref{tmununcft} comes from the finite counterterm of the holographic renormalization; we fix it as in \eqref{kappastu}.

We can not solve the equations \eqref{pw1e}-\eqref{pw5e} analytically; adapting numerical techniques
developed in \cite{Aharony:2007vg}, we solve these equations (subject to the asymptotics
\eqref{uvpw} and \eqref{irpw}) numerically. The results of numerical analysis are data
files assembled of parameters \eqref{pwcond}, labeled by $b$ and $\eta$.  
As for the $\cftstu$ model, we validate the numerical data 
verifying the differential constraint from the first law of the thermodynamics $d\cale=Tds$
(FL)  and the algebraic constraint from the thermodynamic relation $\calf=-P_L$ (TR):
\begin{equation}
\begin{split}
{\rm FL:}\ 0=&\frac{4\sqrt{3} a_{2,h,0} c_{h,0}^2 r_{h,0}^2}{a_{1,h,0}(3 a_{2,h,0}-4 b a_{2,h,0}')} \biggl(
8 (4 b \eta^2+3 \alpha_{1,0}) (\alpha_{1,0}-\alpha_{1,0}' b)-4 \sqrt{b} (2 \chi_{1,0}' b
-3 \chi_{1,0}) \eta\\
&-3 a_{4,2,0}' b-8 a_{2,2,0}' b-32 b^2+6 a_{4,2,0}+16 a_{2,2,0}+3\biggr)
\biggl( a_{2,h,0}^4 (16 c_{h,0}^2 (r_{h,0}^6+c_{h,0}^2)\\
&-r_{h,0}^6 ((c_{h,0}^4-1)^2 r_{h,0}^6-16 c_{h,0}^6))+512 b^2 c_{h,0}^4 r_{h,0}^8
\biggr)^{-1/2}-1\,,
\end{split}
\eqlabel{flpw}
\end{equation}
\begin{equation}
\begin{split}
{\rm TR:}\ 0=&\frac{4\sqrt{3} r_{h,0}^2 c_{h,0}^2}{9a_{1,h,0}} \biggl(
64 b^2 \eta^4+96 \alpha_{1,0} b \eta^2+72 \sqrt{b} \eta \chi_{1,0}+72 \alpha_{1,0}^2-384 b^2+36 a_{2,2,0}\\
&+18 a_{4,2,0}
+9\biggr)
\biggl( a_{2,h,0}^4 (16 c_{h,0}^2 (r_{h,0}^6+c_{h,0}^2)-r_{h,0}^6 ((c_{h,0}^4-1)^2 r_{h,0}^6-16 c_{h,0}^6))
\\&+512 b^2 c_{h,0}^4 r_{h,0}^8\biggr)^{-1/2}-1\,.
\end{split}
\eqlabel{thpw}
\end{equation}

In appendix \ref{errorncft} we have verified FT and TR in the $\ncft$ model with $m/\sqrt{2B}=1$
numerically. 

Technical details presented here are enough to generate $\ncft$ model plots reported in
section \ref{intro}.

\subsubsection{$\cftpw$}\label{m=0}

The $\cftpw$ model is a special case of the $\ncft$ model when the hypermultiplet mass
$m$ is set to zero. This necessitates setting the non-normalizable coefficients 
$\alpha_{1,1}$ and $\chi_{0}$ to zero $\Longrightarrow\ \eta=0$ in \eqref{defeta}.
From \eqref{pw5e} it is clear that this $m=0$ limit is consistent with
\begin{equation}
\eta(x)\equiv 0\qquad \Longrightarrow\qquad \chi_{1,0}=0\,,
\eqlabel{killeta}
\end{equation}
implying that the $\zet_2$ symmetry of the holographic dual, \ie the symmetry associated with
$\chi\leftrightarrow -\chi$, is unbroken. 
In what follows, we study the $\zet_2$-symmetric phase
of the $\cftpw$ anisotropic
thermodynamics\footnote{It is interesting to investigate whether this $\zet_2$ symmetry
can be spontaneously broken, and if so, what is the role of the magnetic field. This,
however, is outside the scope of the current paper.},
\begin{equation}
\calo_3=0\,.
\eqlabel{o30}
\end{equation}

In appendix \ref{ctfpwp} we verified FT and TR in $\cftpw$ to order $\calo(b^4)$
inclusive; we also present $\calo(B^4/T^8)$ results for $R_\cftpw$ and confirm that the
renormalization scheme choice of $\kappa$ as in \eqref{kappastu} leads to
\begin{equation}
R_{\cftpw}=R_{\cftstu}+\calo\left(\frac{B^4}{T^{8}}\right) \,.
\eqlabel{comparepwstu}
\end{equation}

\subsubsection{$\cftminfty$}\label{minfty}

The holographic dual to the  $\cftminfty$ model can be obtained as a particular decoupling limit
$\chi\to \infty$  of the effective action \eqref{pw1}.
As emphasized originally in \cite{HoyosBadajoz:2010td}, the supersymmetric vacuum,
and the isotropic thermal equilibrium states of the theory
\cite{Buchel:2019qcq,Buchel:2007mf} are locally
that of the $4+1$ dimensional conformal plasma. We derive the $5+1$ dimensional holographic
effective action $S_\cftminfty$ (trivially) generalizing the arguments of \cite{HoyosBadajoz:2010td}.

It is the easiest to start with the $\caln=2^*$ vacuum in a holographic dual, the PW geometry \cite{Pilch:2000ue}.
The IR limit corresponds to $\chi\to \infty$, thus, introducing a new radial coordinate $u\to\infty$,
\begin{equation}
e^{2\chi}\simeq 2u\,,\qquad e^{6\alpha}\simeq \frac{2}{3u}\,,\qquad e^A\simeq\left(\frac{2}{3u^4}\right)^{1/3} {\rm k}\,,
\eqlabel{change}
\end{equation}
the background metric becomes 
\begin{equation}
ds_{PW}^2\simeq\left(\frac{3}{2u^2}\right)^{4/3}\left[4 du^2 +\left(\frac{2{\rm k}}{3}\right)^2 \eta_{\mu\nu}dx^\mu dx^\nu\right]\,.
\eqlabel{irmetric}
\end{equation}
The parameter ${\rm k}=2m$ here is defined as in PW \cite{Pilch:2000ue,Buchel:2000cn}.
Introducing \cite{HoyosBadajoz:2010td}
\begin{equation}
e^{4\phi_2}\equiv e^{2(\alpha-\chi)}\simeq \left(\frac{1}{12u^4}\right)^{1/3}\,,\qquad e^{4\phi_1}\equiv e^{6\alpha+2\chi}\simeq \frac 43\,,
\eqlabel{phi12}
\end{equation}
the metric \eqref{irmetric} can be understood as a KK reduction of the locally $AdS_6$ metric
on a compact $x_6\sim x_6 +L_6$:
\begin{equation}
ds_6^2=e^{-2\phi_2} ds_{PW}^2+e^{6\phi_2} dx_6^2\,\,\simeq\,\, \frac{3^{3/2}}{2u^2}
\left[4 du^2+\left(\frac{2{\rm k}}{3}\right)^2\eta_{\mu\nu}dx^\mu dx^\nu+\frac 19 dx_6^2\right]\,.
\eqlabel{6metric1}
\end{equation}
The metric \eqref{6metric1} and the scalar $\phi_1$ \eqref{phi12} is a solution \cite{HoyosBadajoz:2010td}
to $d=6$ $\caln=(1,1)$  $F(4)$ SUGRA \cite{Romans:1985tw}
\begin{equation}
S_{F(4)}=\frac{1}{16\pi G_6}\int_{\calm_6} d\xi^6 \sqrt{-g_6} \left(R_6-4(\del\phi_1)^2+e^{-2\phi_1}+e^{2\phi_1}-\frac{1}{ 16}e^{6\phi_1}\right)\,,
\eqlabel{action6}
\end{equation}
where, using the PW five-dimensional Newton's constant $G_5$,
\begin{equation}
\frac{L_6}{G_6}=\frac{1}{G_5}\,.
\eqlabel{defg6}
\end{equation}
Notice that the bulk gauge field in \eqref{pw1} can be reinterpreted as a gauge field in the six-dimensional
metric \eqref{6metric1}
\begin{equation}
\underbrace{\sqrt{-g_{PW}}\ \rho^4 F_{\mu\nu} F^{\mu\nu}}_{{\rm in}\ ds^2_{PW}}\qquad =\qquad
\underbrace{\sqrt{-g_{6}}\ e^{2\phi_1} F_{[6]\mu\nu} F_{[6]}^{\mu\nu}}_{{\rm in}\ ds^2_{6}}\,,
\eqlabel{f26dim}
\end{equation}
leading to
\begin{equation}
\begin{split}
S_{\cftminfty}=\frac{1}{16\pi G_6}\int_{\calm_6} d\xi^6 \sqrt{-g_6} &\biggl(R_6-
4(\del\phi_1)^2+e^{-2\phi_1}+e^{2\phi_1}-\frac{1}{ 16}e^{6\phi_1}\\
&-4  e^{2\phi_1} F_{[6]\mu\nu}
F_{[6]}^{\mu\nu}\biggr)\,,
\end{split}
\eqlabel{actionminfty}
\end{equation}
which is precisely the (truncated) effective action of the $F(4)$ gauged supergravity of
\cite{Chen:2019qib}\footnote{The identification is as follows:
$A^i=0$, $B=0$, $X=e^{-\phi_1}$, $m=\frac 14$ and  $g^2=\frac 12$.}.

Solutions to the gravitational theory \eqref{actionminfty} representing magnetic branes
dual to anisotropic magnetized $\cftminfty$ plasma correspond to the following background
ansatz:
\begin{equation}
ds_5^2=-c_1^2\ dt^2+ c_2^2\ \left(d\hx^2+d\hat{y}^2\right)+ c_3^3\left(d\hat{z}^2+
d\hx_6\right)+c_4^2\ dr^2\,,\qquad F_{[6]}=B_{[6]}\ d\hx\wedge d\hy\,,
\eqlabel{backminfty}
\end{equation}
where all the metric warp factors $c_i$ as well as the bulk scalar $\phi_1$ 
are functions of the radial coordinate $r$. 
The rescaled, \ie $\hat{\ }$ coordinates, are related to PW coordinates $x^\mu$ and
the KK direction $x_6$ as follows (compare with \eqref{6metric1}):
\begin{equation}
\{{\hat t},{\hat{\boldsymbol{x}}}\}\equiv {\hat x}^\mu=\frac {2{\rm k}}{3} x^\mu\,,\qquad {\hat x}_6=\frac 13 x_6\,.
\eqlabel{rescale}
\end{equation}
It is convenient to fix the radial coordinate $r$ and redefine the metric warp factor, the bulk scalar,
and the magnetic field as 
\begin{equation}
\begin{split}
&c_1=\frac{3^{3/4}r}{2^{1/2}}\ \left(1-\frac{r_0^5}{r^5}\right)^{1/2}\ a_1\,,\qquad
c_2=\frac{3^{3/4}r}{2^{1/2}}\ \ a_2\,,\qquad c_3=\frac{3^{3/4}r}{2^{1/2}}\,,\\
&c_4=\frac{3^{3/4} 2^{1/2}}{r}\ \left(1-\frac{r_0^5}{r^5}\right)^{-1/2}\ a_4\,,\qquad
B_{[6]}=\frac 12 r_0^2\ \hb\,,\qquad \phi_1=\frac 14 \ln\frac 43+p \,.
\end{split}
\eqlabel{redefminfty}
\end{equation}
The radial coordinate $r$ changes 
\begin{equation}
r\ \in\ [r_0,+\infty)\,,
\eqlabel{rrange2}
\end{equation}
where $r_0$ is a location of a regular Schwarzschild horizon, and $r\to +\infty$ is the asymptotic
$AdS_6$ boundary\footnote{
$AdS_6$ of radius $L_{AdS_6}=3^{3/4} 2^{1/2}$ is a solution with $r_0=0$, $\hb=0$ and $a_1=a_2=a_4\equiv 1$
and $p\equiv 0$.}. The bulk scalar field $p$ is dual to a dimension $\Delta=3$ of the effective
five-dimensional boundary  conformal theory. Introducing a radial coordinate $x$ as in \eqref{defx}
we obtain the following system of ODEs (in a radial coordinate $x$, $'=\frac{d}{dx}$):
\begin{equation}
\begin{split}
&0=a_1'+\frac{a_1}{36a_2^3 x (x^5-1) (2 a_2-a_2' x)} \biggl(
18 x^2 a_2^2 (x^5-1) (2 a_2^2 (p')^2-(a_2')^2)\\
&+18 x a_2^3 (3 x^5-8) a_2'-4 a_4^2 (27 a_2^4-8 \hb^2 x^4) e^{2p}-9 a_2^4 (9 a_4^2 e^{-2 p}-e^{6p} a_4^2-20)
\biggr)\,,
\end{split}
\eqlabel{mi1}
\end{equation}
\begin{equation}
\begin{split}
&0=a_4'-\frac{a_4}{36 a_2^4 x (x^5-1) (2 a_2-a_2' x)} \biggl(
18 a_2^3 x^2 (1-x^5) (2 a_2^2 (p')^2+3 (a_2')^2)\\
&+x (90 a_2^4 (x^5-2)+a_4^2 (32 e^{2p} \hb^2 x^4+9 a_2^4 (12 e^{2p}+9 e^{-2 p}-e^{6p}))) a_2'\\
&-3 a_2 (a_4^2 (32 e^{2p} \hb^2 x^4+3 a_2^4 (12 e^{2p}+9 e^{-2 p}-e^{6p}))-60 a_2^4)
\biggr)\,,
\end{split}
\eqlabel{mi2}
\end{equation}
\begin{equation}
\begin{split}
&0=a_2''-\frac{(a_2')^2}{a_2}+\frac{1}{36(x^5-1) x a_2^4} \biggl(a_4^2 (32 e^{2p} \hb^2 x^4
+9 a_2^4 (12 e^{2p}+9 e^{-2 p}-e^{6p}))\\
&+36 a_2^4 (x^5-1)\biggr) a_2'-\frac{32e^{2p} a_4^2 \hb^2 x^2}{9a_2^3 (x^5-1)}\,,
\end{split}
\eqlabel{mi3}
\end{equation}
\begin{equation}
\begin{split}
&0=p''+\frac{1}{36(x^5-1) x a_2^4} \biggl(
a_4^2 (32 e^{2p} \hb^2 x^4+9 a_2^4 (12 e^{2p}+9 e^{-2 p}-e^{6p}))
+36 a_2^4 (x^5-1)\biggr) p'\\
&+\frac{a_4^2}{36a_2^4 x^2 (x^5-1)}
\biggl(32 e^{2p} \hb^2 x^4-27 a_2^4 (4 e^{2p}-3 e^{-2 p}-e^{6p})\biggr)\,.
\end{split}
\eqlabel{mi4}
\end{equation}
As before, $r_0$ is completely scaled out of all the equations of motion.
Eqs.~\eqref{mi1}-\eqref{mi4} have to be solved subject to the following asymptotics:
\nxt in the UV, \ie as $x\to 0_+$,
\begin{equation}
\begin{split}
a_1=&1+a_{1,5} x^5+\calo(x^9)\,,\qquad a_2=1+\frac89 \hb^2 x^4+a_{2,5} x^5+\calo(x^7)\,,\\
a_4=&1-\frac43 \hb^2 x^4-(a_{1,5}+2 a_{2,5}) x^5+\calo(x^6)\,,\qquad p=p_3 x^3+\frac 49 \hb^2 x^4+\calo(x^6)\,;
\end{split}
\eqlabel{uvpwinfty}
\end{equation}
\nxt in the IR, \ie as $y\equiv 1-x\to 0_+$,
\begin{equation}
\begin{split}
&a_1=a_{1,h,0}+\calo(y)\,,\qquad a_2=a_{2,h,0}+\calo(y)\,,\qquad p=\ln p_{h,0}+\calo(y)\,,\\
&a_4=\frac{30 a_{2,h,0}^2p_{h,0}}{(5p_{h,0}^4(9a_{2,h,0}^4(12-p_{h,0}^4)+32\hb^2)+405a_{2,h,0}^4)^{1/2}}+\calo(y)\,.
\end{split}
\eqlabel{irpwinfty}
\end{equation}
In total, given $\hb$, the asymptotic expansions are specified by 6 parameters:
\begin{equation}
\{a_{1,5}\,,\ a_{2,5}\,,\ p_3\,,\ a_{1,h,0}\,,\ a_{2,h,0}\,,\ p_{h,0}\}\,,
\eqlabel{pinftycond}
\end{equation}
which is the correct number of parameters necessary to provide a solution to a system of two
second order and two first order equations, $2\times 2+2\times 1=6$. The parameter 
$p_3$ corresponds to the expectation value of a dimension $\Delta=3$
operator of the boundary theory; the other two parameters, $a_{1,5}$ and $a_{2,5}$, determine
the expectation value of its stress-energy tensor. Using the standard holographic renormalization
we find:
\begin{equation}
\begin{split}
&\langle T_{[5]\hat{t}\hat{t}}\rangle\equiv \cale_{[5]}=\frac{27r_0^5}{32\pi G_6}  (1-2 a_{1,5}+a_{2,5})\,,
\\
&\langle T_{[5]\hx\hx}\rangle=\langle T_{[5]\hy\hy}\rangle\equiv P_{[5]T}=\frac{27r_0^5}{128\pi G_6}
(1-2 a_{1,5}+6 a_{2,5})\,,
\\
&\langle T_{\hz\hz}\rangle=\langle T_{[5]\hx_6\hx_6}\rangle\equiv P_{[5]L}=\frac{27r_0^5}{128\pi G_6}
(1-2 a_{1,5}-4 a_{2,5})\,,
\end{split}
\eqlabel{tmunupinfty}
\end{equation}
for the components of the boundary stress-energy tensor, and
\begin{equation}
s_{[5]}=\frac{27r_0^4 a_{2,h,0}^2}{16G_6}\,,\qquad
T_{[5]}=\frac{\sqrt{5}r_0 a_{1,h,0}}{48\pi a_{2,h,0}^2 p_{h,0}} \biggl[9 a_{2,h,0}^4 (9-p_{h,0}^8+12 p_{h,0}^4)+32
\hb^2 p_{h,0}^4\biggr]^{1/2}\,,
\eqlabel{stpinfty}
\end{equation}
for the entropy density and the temperature.
Note that, 
\begin{equation}
\langle T^\mu_{[5]\ \mu}\rangle=0\,.
\eqlabel{tracepinfty}
\end{equation}
There is no renormalization scheme dependence in \eqref{tmunupinfty}, and the
trace of the stress-energy tensor vanishes --- there is no invariant dimension-five operator
that can be constructed only with the magnetic field strength. 
The (holographic) free energy density is given by the
standard relation \eqref{fstu}. In \eqref{tmunupinfty}-\eqref{stpinfty}
we used the subscript $\ _{[5]}$ to indicate that the thermodynamic quantities are measured
from the perspective of the effective five-dimensional boundary conformal theory; 
to convert to the four-dimensional perspective, we need to account for \eqref{rescale},
see also \cite{Buchel:2019qcq},
\begin{equation}
\begin{split}
&\biggl\{\cale,P_T,P_L\biggr\}=\biggl\{\cale_{[5]},P_{[5]T},P_{[5]L}\biggr\}\ \times\
\underbrace{\left(\frac{2{\rm k}}{3}\right)^4}_{(d\hat{t}\cdot d\hat{vol}_{3})/(dt\cdot d{vol}_{3})}\ \times\
\underbrace{\frac{L_6}{3}}_{\oint d\hx_6}\,, \\
&s=s_{[5]} \ \times\
\underbrace{\left(\frac{2{\rm k}}{3}\right)^3}_{d\hat{vol}_{3}/d{vol}_{3}}\ \times\
\underbrace{\frac{L_6}{3}}_{\oint d\hx_6}\,,\qquad T=T_{[5]}\ \times
\underbrace{\left(\frac{2 {\rm k}}{3}\right)}_{d\hat{t}/dt}\,,\qquad b=\hb\ \times
\underbrace{\left(\frac{2 {\rm k}}{3}\right)^2}_{d\hat{x}\wedge d\hat{y}/dx\wedge dy}\,.
\end{split}
\eqlabel{convertpw}
\end{equation}

As for the other models discussed in this paper, the first law of thermodynamics
$d\cale=T ds$ (at fixed magnetic field) and the thermodynamic relation
$\calf=-P_L$ lead to  constraints on the numerically obtained parameter set \eqref{pinftycond}
(here $'=\frac{d}{d\hb}$):
\begin{equation}
\begin{split}
{\rm FL:}\ 0=&
\frac{6(2 \hb a_{2,5}'-4 \hb a_{1,5}'-5 a_{2,5}+10 a_{1,5}-5) \sqrt{5} a_{2,h,0} p_{h,0}}
{5a_{1,h,0} (32 \hb^2 p_{h,0}^4-9 a_{2,h,0}^4 (p_{h,0}^8-12 p_{h,0}^4-9))^{1/2} (a_{2,h,0}' \hb-a_{2,h,0})}-1\,,
\end{split}
\eqlabel{flpinfty}
\end{equation}
\begin{equation}
\begin{split}
{\rm TR:}\ 0=&\frac{6 (1-2 a_{1,5}) p_{h,0} \sqrt{5}}{
a_{1,h,0} (32 \hb^2 p_{h,0}^4-9 a_{2,h,0}^4
(p_{h,0}^8-12 p_{h,0}^4-9))^{1/2}}-1\,.
\end{split}
\eqlabel{trpinfty}
\end{equation}
In appendix \ref{ctfminftyp} we verified FT and TR in the $\cftminfty$ model to order $\calo(\hb^4)$
inclusive; we also present $\calo(B^4/T^8)$ results for $R_{\cftminfty}$.

\section*{Acknowledgments}
Research at Perimeter
Institute is supported by the Government of Canada through Industry
Canada and by the Province of Ontario through the Ministry of
Research \& Innovation. This work was further supported by
NSERC through the Discovery Grants program.

\appendix

\section{Proof of $-P_L=\cale - sT$ in holographic magnetized plasma}\label{proof}

The proof follows the argument for the universality of the shear viscosity to the entropy
density in holographic plasma  \cite{Buchel:2003tz}.

Consider a  holographic dual to a four dimensional\footnote{Generalization
to other dimensions is straightforward.} gauge theory in an external magnetic field. We are going to assume that the magnetic field is along the $z$-direction, as in \eqref{backstu}.
We take the (dimensionally reduced --- again, this can be relaxed)
holographic background geometry to be
\begin{equation}
ds_5^2=-c_1^2\ dt^2+c_2^2\ \left(dx^2+dy^2\right)+c_3^2\ dz^2+c_4^2\ dr^2\,,\qquad c_i=c_i(r)\,.
\eqlabel{pr1}
\end{equation}
At extremality (whether or not the extremal solution is singular or not within the
truncation is irrelevant), the Poincare symmetry of the background geometry guarantees that
\begin{equation}
R_{tt}+R_{zz}=0\,,
\eqlabel{pr2}
\end{equation}
where $R_{\mu\nu}$ is the Ricci tensor in the orthonormal frame. Clearly, an analogous
condition must be satisfied for the full gravitational stress tensor of the matter supporting
the geometry
\begin{equation}
T_{tt}+T_{zz}=0\,.
\eqlabel{pr3}
\end{equation}
Because turning on the nonextremality will not modify \eqref{pr3}, we see that \eqref{pr2}
is valid away from extremality as well. Computing the Ricci tensor for
\eqref{pr1} reduces \eqref{pr2} to
\begin{equation}
0=R_{tt}+R_{zz}=\frac{1}{c_1c_2^2c_3c_4}\ \frac{d}{dr}\biggl[\left(\frac{c_1}{c_3}\right)'\
\frac{c_2^2c_3^2}{c_4}\biggr] \qquad \Longrightarrow\qquad \left(\frac{c_1}{c_3}\right)'\
\frac{c_2^2c_3^2}{c_4}={\rm const}\,.
\eqlabel{pr4}
\end{equation}
Explicitly evaluating the ratio of the const in \eqref{pr4} in the UV ($r\to \infty$)
and IR ($r\to r_{horizon}$) we recover
\begin{equation}
0=\frac{\cale+P_L}{sT}-1\,,
\eqlabel{pr5}
\end{equation}
for each of the models we study.

We should emphasize that the condition \eqref{pr2} can be explicitly verified using the equations
of motion in each model studied. The point of the argument above (as the related
one in \cite{Buchel:2003tz}) is that this relation is
true based on the symmetries of the problem alone.

\section{Conformal models in the limit $T/\sqrt{B}\gg 1$}\label{highT}

In holographic models, supersymmetry at extremality typically guarantees that equilibrium
isotropic thermodynamics is renormalization scheme independent (compare the $\caln=2^*$ model with the
same masses for the bosonic and the fermionic components $m_b^2=m_f^2$, versus the same model
with $m_b^2\ne m_f^2$ \cite{Buchel:2007vy}). This is not the case for the
holographic magnetized gauge theory plasma in four space-time
dimensions, \eg see
\cite{Endrodi:2018ikq} for $\caln=4$ SYM. In this appendix we discuss the high temperature
anisotropic equilibrium thermodynamics of the conformal (supersymmetric in vacuum) models.
For the (locally) four dimensional models ( $\cftd$, $\cftstu$ and $\cftpw$ ) matching
high-temperature equations of state is a natural way to relate renormalization
schemes in various theories. In the $\cftminfty$ model, which is locally five dimensional,
magnetized thermodynamics is scheme independent.

\subsection{$\cftstu$}\label{bstu}

The high temperature expansion corresponds to the perturbative expansion in $b$.
In what follows we study anisotropic thermodynamics to order $\calo(b^4)$ inclusive.
Introducing
\begin{equation}
\begin{split}
&a_1=1+\sum_{n=1}^\infty a_{1,(n)}\ b^{2n}\,,\qquad a_2=1+\sum_{n=1}^\infty a_{2,(n)}\ b^{2n}\,,\qquad a_4=1+\sum_{n=1}^\infty a_{4,(n)}\ b^{2n}\,,\\
&\rho=1+\sum_{n=1}^\infty \rho_{(n)}\ b^{2n}\,,\qquad \nu=1+\sum_{n=1}^\infty \nu_{(n)}\ b^{2n}\,,
\end{split}
\eqlabel{pert1stu}
\end{equation}
so that (see \eqref{uvstu} and \eqref{irstu} for the asymptotics)
\begin{equation}
\begin{split}
&a_{1,2}=\sum_{n=1}^\infty a_{1,2,(n)}\ b^{2n}\,,\qquad a_{2,2}=\sum_{n=1}^\infty a_{2,2,(n)}\ b^{2n}\,,\qquad
r_{1}=\sum_{n=1}^\infty r_{1,(n)}\ b^{2n}\,,\\
&n_{1}=\sum_{n=1}^\infty n_{1,(n)}\ b^{2n}\,,\qquad a_{1,h,0}=1+\sum_{n=1}^\infty a_{1,h,0,(n)}\ b^{2n}\,,
\qquad  a_{2,h,0}=1+\sum_{n=1}^\infty
a_{2,h,0,(n)}\ b^{2n}\,,\\
&r_{h,0}=1+\sum_{n=1}^\infty r_{h,0,(n)}\ b^{2n}\,,\qquad n_{h,0}=1+\sum_{n=1}^\infty n_{h,0,(n)}\ b^{2n}\,,\  
\end{split}
\eqlabel{vevsstu}
\end{equation}
we find
\nxt at order $n=1$:
\begin{equation}
\begin{split}
0=&a_{2,(1)}''+\frac{x^4+3}{x (x^4-1)} a_{2,(1)}'-\frac{128 x^2}{x^4-1}\,,
\end{split}
\eqlabel{stun11}
\end{equation}
\begin{equation}
\begin{split}
0=&a_{4,(1)}'-\frac{4x^4}{3(x^4-1)} a_{2,(1)}'+\frac{4 (16 x^4+a_{4,(1)})}{x (x^4-1)}\,,
\end{split}
\eqlabel{stun12}
\end{equation}
\begin{equation}
\begin{split}
0=&a_{1,(1)}'+\frac{2(x^4-3)}{3(x^4-1)} a_{2,(1)}'+\frac{4(16 x^4-3 a_{4,(1)})}{3x (x^4-1)}\,,
\end{split}
\eqlabel{stun13}
\end{equation}
\begin{equation}
\begin{split}
0=&\rho_{(1)}''+\frac{x^4+3}{x (x^4-1)} \rho_{(1)}'+\frac{4(16 x^4-3 \rho_{(1)})}{3x^2 (x^4-1)}\,,
\end{split}
\eqlabel{stun14}
\end{equation}
\begin{equation}
\begin{split}
0=&\nu_{(1)}''+\frac{x^4+3}{x (x^4-1)} \nu_{(1)}'-\frac{4 (16 x^4+\nu_{(1)})}{x^2 (x^4-1)}\,;
\end{split}
\eqlabel{stun15}
\end{equation}
\nxt and  at order $n=2$ (we will not need $\rho_{(2)}$ and $\nu_{(2)}$):
\begin{equation}
\begin{split}
0=&a_{2,(2)}''+\frac{x^4+3}{x (x^4-1)} a_{2,(2)}'-(a_{2,(1)}')^2+\frac{128 x^4+24 a_{4,(1)}}{3x (x^4-1)}
a_{2,(1)}'+\frac{512 x^2 (\nu_{(1)}-\rho_{(1)})}{x^4-1}\\
&-\frac{128 x^2 (2 a_{4,(1)}-3 a_{2,(1)})}{x^4-1}\,,
\end{split}
\eqlabel{stun21}
\end{equation}
\begin{equation}
\begin{split}
0=&a_{4,(2)}'-\frac{4x^4}{3(x^4-1)} a_{2,(2)}'+\frac{4 a_{4,(2)}}{x (x^4-1)}
+2 x \left((\rho_{(1)}')^2+\frac13 (\nu_{(1)}')^2\right)
+\frac{x (x^4-9)}{9(x^4-1)} (a_{2,(1)}')^2\\
&-\frac{4(3 x^4 a_{4,(1)}-3 a_{2,(1)} x^4-32 x^4+6 a_{4,(1)})}{9(x^4-1)} a_{2,(1)}'
+\frac{8((\nu_{(1)})^2+3 (\rho_{(1)})^2)}{3x (x^4-1)}
\\&-\frac{256 x^3 (\nu_{(1)}-\rho_{(1)})}{x^4-1}+\frac{2 (96 x^4 a_{4,(1)}-128 a_{2,(1)} x^4+3 (a_{4,(1)})^2)}
{x (x^4-1)}\,,
\end{split}
\eqlabel{stun22}
\end{equation}
\begin{equation}
\begin{split}
&0=a_{1,(2)}'+\frac{2(x^4-3)}{3(x^4-1)} a_{2,(2)}'-\frac{4}{x (x^4-1)} a_{4,(2)}
+\frac{x (x^4-9)}{9(x^4-1)} (a_{2,(1)}')^2-\frac{8(\nu_{(1)}^2+3 \rho_{(1)}^2)}{3x (x^4-1)}\\
&-\frac{2(3 a_{2,(1)} x^4-3 a_{1,(1)} x^4-64 x^4-9 a_{2,(1)}+12 a_{4,(1)}+9 a_{1,(1)})}{9(x^4-1)}
a_{2,(1)}'-\frac{2}{x (x^4-1)} (a_{4,(1)})^2\\
&+2 x \left((\rho_{(1)}')^2+\frac13 (\nu_{(1)}')^2\right)
+\frac{4(32 x^4-3 a_{1,(1)})}{3x (x^4-1)} a_{4,(1)}
+\frac{64x^3 (a_{1,(1)}-4 a_{2,(1)})}{3(x^4-1)}\\&-\frac{256x^3 (\nu_{(1)}-\rho_{(1)})}{3(x^4-1)}\,.
\end{split}
\eqlabel{stun23}
\end{equation}
Eqs.~\eqref{stun11} and \eqref{stun12} can be solved analytically:
\begin{equation}
\begin{split}
a_{2,(1)}=&32 \biggl(\ln(x) \ln(1+x)-{\rm dilog}(x)+\ln(x) \ln(1+x^2)+{\rm dilog}(1+x)
\\
&+\frac 12 {\rm dilog}(1+x^2)\biggr)+\frac{16}{3} \pi^2\,,\\
a_{4,(1)}=&\frac{16 x^4}{3 (x^4-1)} (\pi^2-8 {\rm dilog}(x)+8 \ln(x) \ln(x^2+1)+4 {\rm dilog}(x^2+1)
+8 {\rm dilog}(1+x)\\
&+8 \ln(x) \ln(1+x)-12 \ln(x))\,,
\end{split}
\eqlabel{stuanal}
\end{equation}
while the remaining ones have to be solved numerically. We find:
\begin{equation}
\begin{tabular}{c |c| c |c|c }
$(n)$ & $a_{1,2,(n)}$ & $a_{2,2,(n)}$& $r_{1,(n)}$ & $n_{1,(n)}$ \\
\hline
$(1)$ & $\frac{16}{3}-\frac{16\pi^2}{9}$& 8 & $-\frac 43\pi^2$   & $4\pi^2$\\
\hline
$(2)$ & 1541.8(0)& -3358.0(0) & & 
\end{tabular}
\eqlabel{resvevstu1}
\end{equation}
\begin{equation}
\begin{tabular}{c |c| c |c|c}
$(n)$ & $a_{1,h,0,(n)}$ & $a_{2,h,0,(n)}$
& $r_{h,0,(n)}$ & $n_{h,0,(n)}$ \\
\hline
$(1)$  & -7.2270(2)& $\frac 43\pi^2$ & -9.770(3)& 29.310(9)
\\
\hline
$(2)$  & 1336.5(8) & -2069.9(8) & & 
\end{tabular}
\eqlabel{resvevstu2}
\end{equation}

An important check on the numerical results are the first law of thermodynamics FL \eqref{flstu}
and the thermodynamic relation TR \eqref{fstu3}. Given the perturbative expansions
\eqref{vevsstu}, we can represent
\begin{equation}
{\rm FL}=\sum_{n=1}^\infty fl_{(n)}\ b^{2n}\,,\qquad {\rm TR}=\sum_{n=1}^\infty tr_{(n)}\ b^{2n}\,,
\eqlabel{fltrstu}
\end{equation}
where 
\nxt at order $n=1$:
\begin{equation}
\begin{split}
&fl_{(1)}:\qquad 0=\frac 23 a_{2,h,0,(1)}-16-a_{1,h,0,(1)}\,,
\\
&tr_{(1)}:\qquad 0=-2 a_{2,h,0,(1)}-\frac{16}{3}-2 a_{1,2,(1)}-a_{1,h,0,(1)}\,;
\end{split}
\eqlabel{rela2stu}
\end{equation}
\nxt and at order $n=2$:
\begin{equation}
\begin{split}
&fl_{(2)}:\qquad 0=\frac{896}{9}-\frac23 a_{1,h,0,(1)} a_{2,h,0,(1)}+a_{1,h,0,(1)}^2
-2 r_{h,0,(1)}^2-\frac 23 n_{h,0,(1)}^2+\frac{19}{9} a_{2,h,0,(1)}^2\\
&\qquad\qquad\qquad +\frac{64}{3} n_{h,0,(1)}
-\frac{64}{3} r_{h,0,(1)}-\frac43 a_{2,2,(2)}+\frac{32}{3} a_{2,h,0,(1)}+\frac{10}{3} a_{2,h,0,(2)}
\\
&\qquad\qquad\qquad+16 a_{1,h,0,(1)}-a_{1,h,0,(2)}+2 a_{1,2,(2)}\,,
\\
&tr_{(2)}:\qquad 0=2 a_{1,h,0,(1)} a_{2,h,0,(1)}+2 a_{1,h,0,(1)} a_{1,2,(1)}+a_{1,h,0,(1)}^2+4 a_{2,h,0,(1)} a_{1,2,(1)}
+\frac{128}{3}
\\
&\qquad\qquad\qquad+3 a_{2,h,0,(1)}^2-2 r_{h,0,(1)}^2-\frac23 n_{h,0,(1)}^2+\frac{32}{3} a_{1,2,(1)}-2 a_{1,2,(2)}
+\frac{16}{3} a_{1,h,0,(1)}\\
&\qquad\qquad\qquad
-a_{1,h,0,(2)}+32 a_{2,h,0,(1)}-2 a_{2,h,0,(2)}-\frac{64}{3} r_{h,0,(1)}
+\frac{64}{3}n_{h,0,(1)}\,.
\end{split}
\eqlabel{rela4stu}
\end{equation}
Using the results \eqref{resvevstu1} and \eqref{resvevstu2} (rather, we use more precise values of the
parameters reported --- obtained from numerics with 40 digit precision) we find
\nxt at order $n=1$:
\begin{equation}
\begin{split}
&fl_{(1)}:\qquad 0=-7.7822(6)\times 10^{-15}\,,\qquad  tr_{(1)}:\qquad 0=-7.1054(3)\times 10^{-15}\,;
\end{split}
\eqlabel{rela2stuf}
\end{equation}
\nxt and at order $n=2$:
\begin{equation}
\begin{split}
&fl_{(2)}:\qquad 0=-1.9681(5)\times 10^{-6}\,,\qquad  tr_{(2)}:\qquad 0=2.4872(6)\times 10^{-6}\,.
\end{split}
\eqlabel{rela4stuf}
\end{equation}

 Using the perturbative expansion \eqref{vevsstu}, it is straightforward to invert the relation  between
 $T/\sqrt{B}$ and $b$ (see \eqref{ststu} and \eqref{redefstu}), and use the results \eqref{tmunustu}
 with \eqref{kappastu}, along with the analytical values for the parameters \eqref{resvevstu1}
 and \eqref{resvevstu2} (and the analytical expression for $a_{1,h,0,(1)}$ obtained from \eqref{rela2stu})
 to arrive at 
\begin{equation}
\begin{split}
&R_\cftstu=1-\frac{4B^2}{\pi^4 T^4}\ \ln\frac{T}{\mu\sqrt{2}}
+\left(
\frac{\pi^2}{18}+\frac{a_{2,2,(2)}}{512}-\frac23+8\ \ln^2 \frac{T}{\mu\sqrt{2}}
\right)\ \frac{B^4}{\pi^8 T^8}+\cdots\\
&=1-\frac{4B^2}{\pi^4 T^4}\ \ln\frac{T}{\mu\sqrt{2}}
+\left(-6.67694906(1)+8\ \ln^2 \frac{T}{\mu\sqrt{2}}
\right)\ \frac{B^4}{\pi^8 T^8}+\calo\left(\frac{B^6}{T^{12}}\ln^3\frac{T}{\mu}\right)\,.
\end{split}
\eqlabel{rstupertfin}
\end{equation}
It is important to keep in mind that the value $a_{2,2,(2)}$ is sensitive to the matter content
of the gravitational dual --- set of relevant operators in $\cftstu$ that develop expectation
values in anisotropic thermal equilibrium.  

\subsection{$\cftpw$}\label{ctfpwp}

The high temperature expansion of the $\zet_2$ symmetric, \ie $\chi\equiv0$ phase, 
of anisotropic $\cftpw$ plasma thermodynamics corresponds to the perturbative expansion in $b$.
In what follows we study anisotropic thermodynamics to order $\calo(b^4)$ inclusive.
Introducing
\begin{equation}
\begin{split}
&a_1=1+\sum_{n=1}^\infty a_{1,(n)}\ b^{2n}\,,\qquad a_2=1+\sum_{n=1}^\infty a_{2,(n)}\ b^{2n}\,,\qquad a_4=1+\sum_{n=1}^\infty a_{4,(n)}\ b^{2n}\,,\\
&\alpha=\sum_{n=1}^\infty \alpha_{(n)}\ b^{2n}\,,
\end{split}
\eqlabel{pert1pw}
\end{equation}
so that (see \eqref{uvpw} and \eqref{irpw} for the asymptotics)
\begin{equation}
\begin{split}
&a_{2,2,0}=\sum_{n=1}^\infty a_{2,2,0,(n)}\ b^{2n}\,,\qquad a_{4,2,0}=\sum_{n=1}^\infty a_{4,2,0,(n)}\ b^{2n}\,,\qquad
\alpha_{1,0}=\sum_{n=1}^\infty \alpha_{1,0,(n)}\ b^{2n}\,,\\
&a_{1,h,0}=1+\sum_{n=1}^\infty a_{1,h,0,(n)}\ b^{2n}\,,
\qquad  a_{2,h,0}=1+\sum_{n=1}^\infty
a_{2,h,0,(n)}\ b^{2n}\,,\\
&r_{h,0}=1+\sum_{n=1}^\infty r_{h,0,(n)}\ b^{2n}\,,
\end{split}
\eqlabel{vevspw}
\end{equation}
we find
\nxt at order $n=1$:
\begin{equation}
\begin{split}
0=&a_{2,(1)}''+\frac{x^4+3}{x (x^4-1)} a_{2,(1)}'-\frac{128 x^2}{x^4-1}\,,
\end{split}
\eqlabel{pwn11}
\end{equation}
\begin{equation}
\begin{split}
0=&a_{4,(1)}'-\frac{4x^4}{3(x^4-1)} a_{2,(1)}'+\frac{4 (16 x^4+a_{4,(1)})}{x (x^4-1)}\,,
\end{split}
\eqlabel{pwn12}
\end{equation}
\begin{equation}
\begin{split}
0=&a_{1,(1)}'+\frac{2(x^4-3)}{3(x^4-1)} a_{2,(1)}'+\frac{4(16 x^4-3 a_{4,(1)})}{3x (x^4-1)}\,,
\end{split}
\eqlabel{pwn13}
\end{equation}
\begin{equation}
\begin{split}
0=&\alpha_{(1)}''+\frac{x^4+3}{x (x^4-1)} \alpha_{(1)}'+\frac{4(16 x^4-3\alpha_{(1)})}{3 x^2(x^4-1)}\,;
\end{split}
\eqlabel{pwn14}
\end{equation}
\nxt and  at order $n=2$ (we will not need $\alpha_{(2)}$):
\begin{equation}
\begin{split}
0=&a_{2,(2)}''+\frac{x^4+3}{x (x^4-1)} a_{2,(2)}'+\frac{128 x^4+24 a_{4,(1)}}{3x (x^4-1)} a_{2,(1)}'
-(a_{2,(1)}')^2\\
&-\frac{128 x^2 (2 a_{4,(1)}+4 \alpha_{(1)}-3 a_{2,(1)})}{x^4-1}\,,
\end{split}
\eqlabel{pwn22}
\end{equation}
\begin{equation}
\begin{split}
&0=a_{4,(2)}'-\frac{4x^4}{3(x^4-1)} a_{2,(2)}'+2 x (\alpha_{(1)}')^2
+\frac{x (x^4-9)}{9(x^4-1)} (a_{2,(1)}')^2
\\&-\frac{4(x^4 (3 a_{4,(1)}-3 a_{2,(1)}-32)+6 a_{4,(1)})}{9(x^4-1)} a_{2,(1)}'
+\frac{8 \alpha_{(1)} (32 x^4+\alpha_{(1)})}{x (x^4-1)}
\\&+\frac{2 (32 x^4 (3 a_{4,(1)}-4 a_{2,(1)})+3 a_{4,(1)}^2+2 a_{4,(2)})}{x (x^4-1)}\,,
\end{split}
\eqlabel{pwn23}
\end{equation}
\begin{equation}
\begin{split}
0=&a_{1,(2)}'+\frac{2(x^4-3)}{3(x^4-1)} a_{2,(2)}'
+\frac{x (x^4-9)}{9(x^4-1)} (a_{2,(1)}')^2+\frac{2}{9(x^4-1)}
\biggl(x^4 (3 a_{1,(1)}-3 a_{2,(1)}+64)\\
&-9 a_{1,(1)}-12 a_{4,(1)}+9 a_{2,(1)}\biggr) a_{2,(1)}'+2 x (\alpha_{(1)}')^2
-\frac{8 \alpha_{(1)}^2}{x (x^4-1)}+\frac{256 x^3 \alpha_{(1)}}{3(x^4-1)}\\
&-\frac{2}{3x (x^4-1)}
\biggl(6 a_{4,(2)}+32 x^4( 4a_{2,(1)}-a_{1,(1)})+2 a_{4,(1)} (-32 x^4+3 a_{1,(1)})+3 a_{4,(1)}^2\biggr)\,.
\end{split}
\eqlabel{pwn21}
\end{equation}
Eqs.~\eqref{pwn11} and \eqref{pwn12} can be solved analytically, see \eqref{stuanal},
while the remaining ones have to be solved numerically. We find:
\begin{equation}
\begin{tabular}{c |c | c |c |c|c |c }
$(n)$ & $a_{2,2,0,(n)}$ & $a_{4,2,0,(n)}$ & $\alpha_{1,0,(n)}$ & $a_{1,h,0,(n)}$ & $a_{2,h,0,(n)}$ 
 & $r_{h,0,(n)}$\\
\hline
$(1)$ &  8  & $\frac{16\pi^2}{9}$  &  $-\frac 43\pi^2$  &-7.2270(2)  & $\frac43\pi^2$ &
-9.770(3)
\\  
\hline
$(2)$ &  -1203.9(2)  &1064.0(4)  &    & 652.34(4) & -863.4(3) &  
\end{tabular}
\eqlabel{resvevpw}
\end{equation}

An important check on the numerical results are the first law of thermodynamics FL \eqref{flpw}
and the thermodynamic relation TR \eqref{thpw}. Given the perturbative expansions
\eqref{vevspw}, and using the representation \eqref{fltrstu}, we find:
\nxt at order $n=1$: 
\begin{equation}
\begin{split}
&fl_{(1)}:\qquad 0=\frac 23 a_{2,h,0,(1)}-16-a_{1,h,0,(1)}\,,
\\
&tr_{(1)}:\qquad 0=-2 a_{2,h,0,(1)}-48-a_{1,h,0,(1)}+4 a_{2,2,0,(1)}+2 a_{4,2,0,(1)}\,;
\end{split}
\eqlabel{rela2pw}
\end{equation}
\nxt and at order $n=2$:
\begin{equation}
\begin{split}
&fl_{(2)}:\ 0=-\frac23 a_{2,h,0,(1)} a_{1,h,0,(1)}+\frac{896}{9}+\frac{32}{3} a_{2,h,0,(1)}
+\frac{10}{3} a_{2,h,0,(2)}-2 r_{h,0,(1)}^2
+\frac{19}{9} a_{2,h,0,(1)}^2\\&+a_{1,h,0,(1)}^2-8 \alpha_{1,0,(1)}^2-a_{1,h,0,(2)}-\frac{16}{3} a_{2,2,0,(2)}
-2 a_{4,2,0,(2)}+16 a_{1,h,0,(1)}-\frac{64}{3} r_{h,0,(1)}\,,\\
&tr_{(2)}:\  0=\frac{2432}{9}-8 a_{2,2,0,(1)} a_{2,h,0,(1)}-4 a_{2,h,0,(1)} a_{4,2,0,(1)}-4 a_{1,h,0,(1)} a_{2,2,0,(1)}
\\&-2 a_{1,h,0,(1)} a_{4,2,0,(1)}+a_{1,h,0,(1)}^2+48 a_{1,h,0,(1)}-a_{1,h,0,(2)}+\frac{352}{3} a_{2,h,0,(1)}-2 a_{2,h,0,(2)}
\\&+2 a_{2,h,0,(1)} a_{1,h,0,(1)}+3 a_{2,h,0,(1)}^2-2 r_{h,0,(1)}^2+8 \alpha_{1,0,(1)}^2-\frac{64}{3} r_{h,0,(1)}
-\frac{64}{3} a_{2,2,0,(1)}\\&+4 a_{2,2,0,(2)}-\frac{32}{3} a_{4,2,0,(1)}+2 a_{4,2,0,(2)}\,.
\end{split}
\eqlabel{rela4pw}
\end{equation}
Using the results \eqref{resvevpw}  (rather, we use more precise values of the
parameters reported --- obtained from numerics with 40 digit precision) we find
\nxt at order $n=1$:
\begin{equation}
\begin{split}
&fl_{(1)}:\qquad 0=-7.7822(6)\times 10^{-15}\,,\qquad  tr_{(1)}:\qquad 0=-2.9555(5)\times 10^{-15}\,;
\end{split}
\eqlabel{rela2pwf}
\end{equation}
\nxt and at order $n=2$:
\begin{equation}
\begin{split}
&fl_{(2)}:\qquad 0=-1.6451(1)\times 10^{-6}\,,\qquad  tr_{(2)}:\qquad 0=2.2505(2)\times 10^{-6}\,.
\end{split}
\eqlabel{rela4pwf}
\end{equation}

 Using the perturbative expansion \eqref{vevspw}, it is straightforward to invert the relation  between
 $T/\sqrt{B}$ and $b$ (see \eqref{stncft} and \eqref{redefstu}), and use the results \eqref{tmununcft}
 with \eqref{kappastu}, along with the analytical values for the parameters \eqref{resvevpw},
 to arrive at 
\begin{equation}
\begin{split}
&R_\cftpw=1-\frac{4B^2}{\pi^4 T^4}\ \ln\frac{T}{\mu\sqrt{2}}
+\left(\frac{\pi^2}{18}+\frac{a_{2,2,0,(2)}}{512}-\frac23+8\ \ln^2 \frac{T}{\mu\sqrt{2}}
\right)\ \frac{B^4}{\pi^8 T^8}+\cdots\\
&=1-\frac{4B^2}{\pi^4 T^4}\ \ln\frac{T}{\mu\sqrt{2}}
+\left(-2.4697(5)+8\ \ln^2 \frac{T}{\mu\sqrt{2}}
\right)\ \frac{B^4}{\pi^8 T^8}+\calo\left(\frac{B^6}{T^{12}}\ln^3\frac{T}{\mu}\right)\,.
\end{split}
\eqlabel{rpwpertfin}
\end{equation}
Note that while the first line in \eqref{rpwpertfin} is equivalent to the corresponding
expression in  \eqref{rstupertfin}, the numerical values (compare the second lines) are different:
this is related to the fact that the value $a_{2,2,0,(2)}$ in the $\cftpw$ dual is ``sourced''
by a single dimension $\Delta=2$ operator (the scalar field $\alpha$ in the holographic dual),
while the value $a_{2,2,(2)}$ in the $\cftstu$ model is ``sourced'' by two dimension
$\Delta=2$ operators (the scalar fields $\rho$ and $\nu$ in the holographic dual).

\subsection{$\cftminfty$}\label{ctfminftyp}

The high temperature expansion corresponds to the perturbative expansion in $\hb$.
In what follows we study anisotropic thermodynamics to order $\calo(\hb^4)$ inclusive.
Introducing
\begin{equation}
\begin{split}
&a_1=1+\sum_{n=1}^\infty a_{1,(n)}\ \hb^{2n}\,,\qquad a_2=1+\sum_{n=1}^\infty a_{2,(n)}\ \hb^{2n}\,,\qquad
a_4=1+\sum_{n=1}^\infty a_{4,(n)}\ \hb^{2n}\,,\\
&p=\sum_{n=1}^\infty p_{(n)}\ \hb^{2n}\,,
\end{split}
\eqlabel{pert1pi}
\end{equation}
so that (see \eqref{uvstu} and \eqref{irstu} for the asymptotics)
\begin{equation}
\begin{split}
&a_{1,5}=\sum_{n=1}^\infty a_{1,5,(n)}\ \hb^{2n}\,,\qquad a_{2,5}=\sum_{n=1}^\infty a_{2,5,(n)}\ \hb^{2n}\,,\qquad
p_{3}=\sum_{n=1}^\infty p_{3,(n)}\ \hb^{2n}\,,\\
&a_{1,h,0}=1+\sum_{n=1}^\infty a_{1,h,0,(n)}\ \hb^{2n}\,,
\qquad  a_{2,h,0}=1+\sum_{n=1}^\infty
a_{2,h,0,(n)}\ \hb^{2n}\,,\\
&p_{h,0}=1+\sum_{n=1}^\infty p_{h,0,(n)}\ \hb^{2n}\,,
\end{split}
\eqlabel{vevspi}
\end{equation}
we find
\nxt at order $n=1$:
\begin{equation}
\begin{split}
0=&a_{2,(1)}''+ \frac{x^5+4}{(x^5-1) x} a_{2,(1)}'-\frac{32x^2}{9(x^5-1)}\,,
\end{split}
\eqlabel{pin11}
\end{equation}
\begin{equation}
\begin{split}
0=&a_{4,(1)}'-\frac{1}{(x^5-1) x}\left(\frac54 a_{2,(1)}' x^6-5 a_{4,(1)}-\frac43 x^4\right)\,,
\end{split}
\eqlabel{pin12}
\end{equation}
\begin{equation}
\begin{split}
0=&a_{1,(1)}'+\frac{3 x^5-8}{4(x^5-1)}a_{2,(1)}'+\frac{1}{(x^5-1) x}\biggl(\frac49 x^4-5 a_{4,(1)}\biggr)\,,
\end{split}
\eqlabel{pin13}
\end{equation}
\begin{equation}
\begin{split}
0=&p_{(1)}''+ \frac{x^5+4}{x (x^5-1)} p_{(1)}'-\frac{1}{x^2 (x^5-1)}\biggl(
6 p_{(1)}-\frac89 x^4\biggr)\,;
\end{split}
\eqlabel{pin14}
\end{equation}
\nxt and  at order $n=2$ (we will not need $p_{(2)}$):
\begin{equation}
\begin{split}
0=&a_{2,(2)}''+ \frac{x^5+4}{x (x^5-1)} a_{2,(2)}'-(a_{2,(1)}')^2+
\frac{2(4 x^4+45 a_{4,(1)})}{9x (x^5-1)} a_{2,(1)}'\\
&+\frac{32x^2}{9(x^5-1)} \biggl(3 a_{2,(1)}-2 a_{4,(1)}-2 p_{(1)}\biggr)\,,
\end{split}
\eqlabel{pin21}
\end{equation}
\begin{equation}
\begin{split}
0=&a_{4,(2)}'-\frac{1}{x (x^5-1)}\biggl(\frac54 a_{2,(2)}' x^6-5 a_{4,(2)}\biggr)
+\frac{x (x^5-6)}{8(x^5-1)} (a_{2,(1)}')^2-\frac{1}{36(x^5-1)} \biggl(
\\&45 a_{4,(1)} x^5-45 a_{2,(1)} x^5-8 x^4+90 a_{4,(1)}\biggr) a_{2,(1)}'+\frac x2 (p_{(1)}')^2
+\frac{1}{6x (x^5-1)} \biggl(
16 x^4 p_{(1)}\\
&+24 x^4 a_{4,(1)}-32 a_{2,(1)} x^4+18 p_{(1)}^2+45 a_{4,(1)}^2\biggr)\,,
\end{split}
\eqlabel{pin22}
\end{equation}
\begin{equation}
\begin{split}
&0=a_{1,(2)}'+\frac{1}{x (x^5-1)}\biggl(\frac34 a_{2,(2)}' x^6-5 a_{4,(2)}-2 a_{2,(2)}' x\biggr)
+\frac{x (x^5-6)}{8(x^5-1)} (a_{2,(1)}')^2+\frac x2 (p_{(1)}')^2
\\&+\frac{1}{36(x^5-1)} \biggl(27 a_{1,(1)} x^5-27 a_{2,(1)} x^5+8 x^4
-72 a_{1,(1)}-90 a_{4,(1)}+72 a_{2,(1)}\biggr) a_{2,(1)}'\\
&-\frac{1}{18x (x^5-1)} \biggl(
45 a_{4,(1)}^2-16 x^4 p_{(1)}-8 a_{1,(1)} x^4-16 x^4 a_{4,(1)}+32 a_{2,(1)} x^4+54 p_{(1)}^2\\
&+90 a_{1,(1)} a_{4,(1)}
 \biggr)\,.
\end{split}
\eqlabel{pin23}
\end{equation}
Eqs.~\eqref{pin11} and \eqref{pin12} can be solved analytically\footnote{However, the resulting expressions
are too long to be presented here. For the same reason we report only the
numerical expression for  $a_{2,h,0,(1)}$.}, while the remaining ones have to be solved numerically. We find:
\begin{equation}
\begin{tabular}{c |c | c |c |c|c |c }
$(n)$ & $a_{1,5,(n)}$ & $a_{2,5,(n)}$ & $p_{3,(n)}$ & $a_{1,h,0,(n)}$ & $a_{2,h,0,(n)}$ 
 & $p_{h,0,(n)}$\\
\hline
$(1)$ & -0.25581(6)   & $-\frac{32}{45}$  & -0.645(2)   & -0.12878(5)  &0.27576(4)  & -0.25155(9)
\\  
\hline
$(2)$ &  0.22327(6)  &  -0.5489(8) &    & 0.20658(5) & -0.2934(9)  &  
\end{tabular}
\eqlabel{resvevpi}
\end{equation}

An important check on the numerical results are the first law of thermodynamics FL \eqref{flpinfty}
and the thermodynamic relation TR \eqref{trpinfty}. Given the perturbative expansions
\eqref{vevspi}, and using the representation \eqref{fltrstu}, we find:
\nxt at order $n=1$:
\begin{equation}
\begin{split}
&fl_{(1)}:\qquad 0=-\frac{4}{45}-\frac25 a_{1,5,(1)}-a_{1,h,0,(1)}+\frac15 a_{2,5,(1)}\,,
\\
&tr_{(1)}:\qquad 0=-\frac{4}{45}-2 a_{2,h,0,(1)}-2 a_{1,5,(1)}-a_{1,h,0,(1)}\,;
\end{split}
\eqlabel{rela2pi}
\end{equation}
\nxt and at order $n=2$:
\begin{equation}
\begin{split}
&fl_{(2)}:\qquad 0=\frac{8}{675}+\frac25 a_{1,5,(1)} a_{1,h,0,(1)}-\frac15 a_{1,h,0,(1)} a_{2,5,(1)}+a_{1,h,0,(1)}^2
+\frac{16}{45} a_{2,h,0,(1)}\\
&\qquad\qquad\qquad+2 a_{2,h,0,(2)}+\frac{4}{45} a_{1,h,0,(1)}-a_{1,h,0,(2)}-\frac{8}{45} p_{h,0,(1)}
-\frac{4}{225} a_{2,5,(1)}\\
&\qquad\qquad\qquad-\frac35 a_{2,5,(2)}+\frac{8}{225} a_{1,5,(1)}+\frac65 a_{1,5,(2)}-\frac35 p_{h,0,(1)}^2+a_{2,h,0,(1)}^2
\,,\\
&tr_{(2)}:\qquad 0=\frac{8}{675}+\frac{8}{15} a_{2,h,0,(1)}-2 a_{2,h,0,(2)}-\frac{8}{45} p_{h,0,(1)}-\frac35 p_{h,0,(1)}^2
+3 a_{2,h,0,(1)}^2\\
&\qquad\qquad\qquad+4 a_{2,h,0,(1)} a_{1,5,(1)}+2 a_{2,h,0,(1)} a_{1,h,0,(1)}+2 a_{1,5,(1)} a_{1,h,0,(1)}+a_{1,h,0,(1)}^2
\\&\qquad\qquad\qquad-2 a_{1,5,(2)}-a_{1,h,0,(2)}+\frac{4}{45} a_{1,h,0,(1)}+\frac{8}{45} a_{1,5,(1)}\,.
\end{split}
\eqlabel{rela4pi}
\end{equation}
Using results \eqref{resvevpi}  (rather, we use more precise values of the
parameters reported --- obtained from numerics with 40 digit precision) we find
\nxt at order $n=1$:
\begin{equation}
\begin{split}
&fl_{(1)}:\qquad 0=1.8010(4)\times 10^{-12}\,,\qquad  tr_{(1)}:\qquad 0=9.8392(9)\times 10^{-12}\,;
\end{split}
\eqlabel{rela2pif}
\end{equation}
\nxt and at order $n=2$:
\begin{equation}
\begin{split}
&fl_{(2)}:\qquad 0=-1.0535(2)\times 10^{-12}\,,\qquad  tr_{(2)}:\qquad 0=-3.3646(2)\times 10^{-12}\,.
\end{split}
\eqlabel{rela4pif}
\end{equation}

 Using the perturbative expansion \eqref{vevspi}, it is straightforward to invert the relation  between
 $T/\sqrt{B}$ and $\hb$ (see \eqref{convertpw}), and arrive at 
\begin{equation}
\begin{split}
R_\cftminfty=&1+\frac{3125}{512} a_{2,5,(1)}\ \frac{B^2}{\pi^4 T^4}+\frac{390625}{4718592} \biggl(90 a_{1,5,(1)} a_{2,5,(1)}
+180 a_{1,h,0,(1)} a_{2,5,(1)}\\
&+180 a_{2,5,(1)}^2+16 a_{2,5,(1)}+45 a_{2,5,(2)}\biggr)\ \frac{B^4}{\pi^8 T^8}+\calo\left(\frac{B^6}{T^{12}}\right)\\
=&1-\frac{625}{144}\ \frac{B^2}{\pi^4 T^4}+7.2682(1)\ \frac{B^4}{\pi^8 T^8}+\calo\left(\frac{B^6}{T^{12}}\right)\,.
\end{split}
\eqlabel{rminftypertfin}
\end{equation}

\section{FT and TR in a $\ncft$ model}\label{errorncft}

\begin{figure}[t]
\begin{center}
\psfrag{x}[cc][][0.7][0]{{${T}/{\sqrt{B}}$}}
\psfrag{y}[cc][][0.7][0]{{${d\cale}/{Tds}-1$}}
\psfrag{z}[cc][][0.7][0]{{${(\cale+P_L)}/{sT}-1$}}
\includegraphics[width=2.8in]{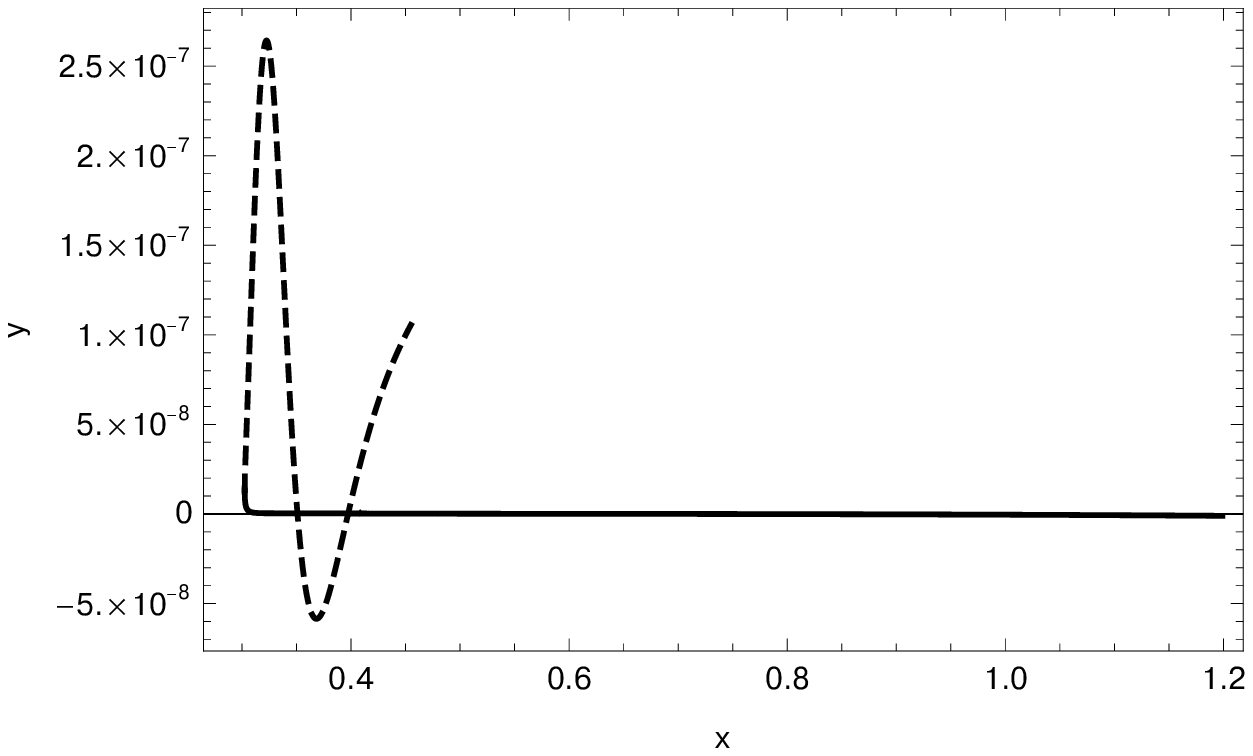}\qquad
\includegraphics[width=2.8in]{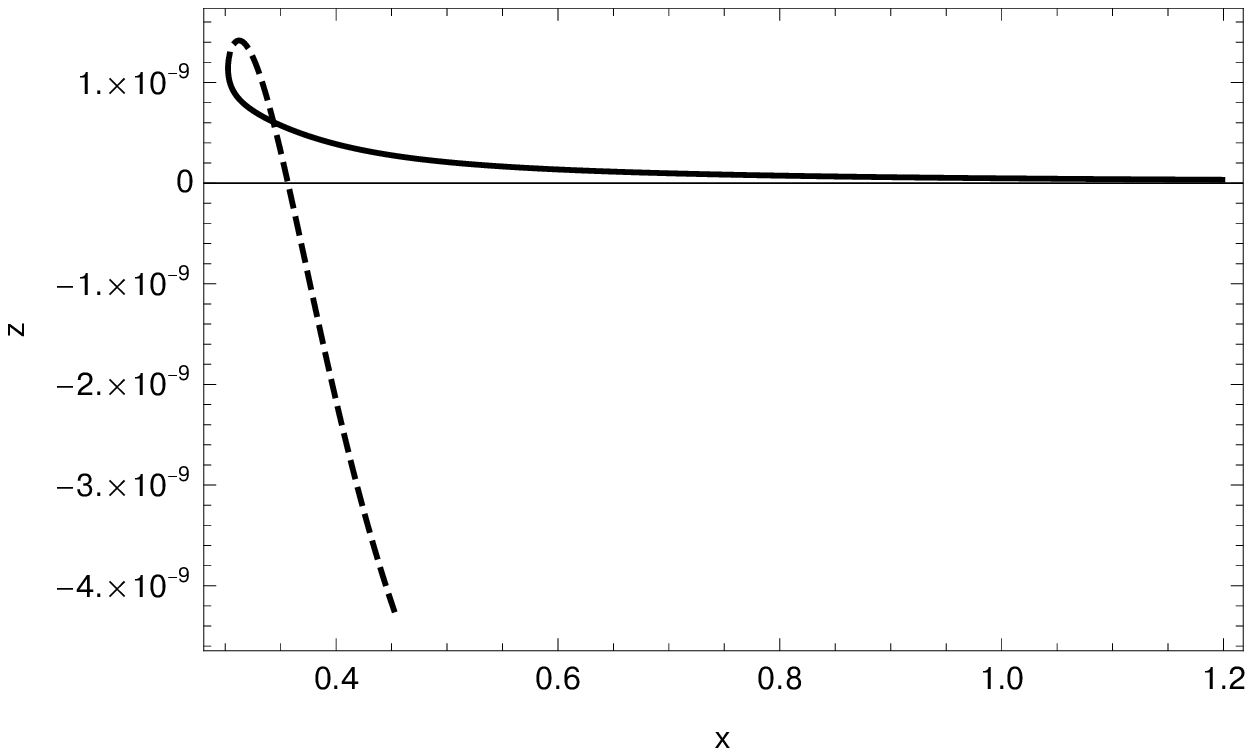}
\end{center}
  \caption{Numerical checks of the first law of thermodynamics $d\cale=Tds$
  (left panel, fixed $B$ and $m$) and the basic thermodynamic relation $\calf=-P_L$
  (right panel) in the $\ncft$ model with $m=\sqrt{2B}$. The dashed parts of the curves
  indicate thermodynamically unstable branches of the model. 
} \label{figuree}
\end{figure}

In appendix \ref{highT} we verified the first law of the thermodynamics (FL) and the
basic thermodynamic relation $\calf=-P_L$ (TR) in various anisotropic magnetized holographic plasma
models perturbatively in $\frac{T}{\sqrt{B}}\gg 1$. In fact, we verified both constraints,
in all the models considered in the paper, for finite values of  $\frac{T}{\sqrt{B}}$.
In Fig.~\ref{figuree} we present the checks on these constraints in the  $\ncft$ model
with $\frac{m}{\sqrt{2B}}=1$.

\bibliographystyle{JHEP}
\bibliography{qgpb2}

\end{document}